\documentclass[twocolumn,groupedaddress,aps,pra,
amsmath,amssymb,showpacs]{revtex4-1}
\usepackage[colorlinks,linkcolor=blue,anchorcolor=blue,
            citecolor=blue,bookmarksnumbered]{hyperref}
\usepackage{graphicx,epsfig,float,epstopdf}
\usepackage{amssymb,amsmath,subeqnarray,mathrsfs,
           amsfonts,bm,makecell,cases,color,extpfeil,lipsum}
\usepackage{dcolumn,comment}\allowdisplaybreaks

\begin{document}
\title{Self-Organized Structures of Two-Component Laser Fields and Their Active Controls in a Cold Rydberg Atomic Gas}
\author{Zeyun Shi$^{1}$ and Guoxiang Huang$^{1,2,3}$}

\affiliation{
$^1$State Key Laboratory of Precision Spectroscopy, East China Normal University, Shanghai 200062, China\\
$^2$NYU-ECNU Joint Institute of Physics, New York University Shanghai, Shanghai 200062, China\\
$^3$Collaborative Innovation Center of Extreme Optics, Shanxi University, Taiyuan, Shanxi 030006, China}

\date{\today}

\begin{abstract}

We investigate the formation and control of stationary optical patterns in a cold Rydberg atomic gas via double electromagnetically induced transparency. We show that, through the modulational instability of plane-wave state of a laser field with two polarization components, the system undergoes a spontaneous symmetry breaking and hence the emergence of plentiful self-organized spatial optical structures, which can be manipulated by the ratio between the cross- and self-Kerr nonlinearities, the nonlocality degree of the Kerr nonlinearities, and the populations initially prepared in the two atomic ground states. Interestingly, a crossover from mixture to separation in space (optical phase separation) of the two polarization components occurs when the ratio between the cross- and self-Kerr nonlinearities exceeds a critical value.
We also show that the system supports nonlocal two-component spatial optical solitons and vortices when the parameters of the system are selected suitably. The rich diversity and active controllability of the self-organized optical structures reported here provide a way for realizing novel optical patterns and solitons and their structural phase transitions based on Rydberg atomic gases.

\pacs{42.65.Sf, 42.65.Tg, 42.50.Gy, 32.80.Ee}


\end{abstract}

\maketitle

\section{Introduction}\label{sec1}

In the past two decades, much attention has been paid to the study of cold Rydberg atomic gases~\cite{Saffman2010,Adams2020} working under condition of electromagnetically induced transparency (EIT)~\cite{Mohapatra2007,Pritchard2010}. EIT is an important quantum interference effect typically occurring in resonant three-level atomic systems, by which the absorption of a probe laser field can be greatly suppressed by a control laser field~\cite{Fleischhauer2005}. Due to the strong interaction between Rydberg atoms (also called Rydberg-Rydberg interaction), Rydberg gases are ideal nonlinear optical media to acquire giant enhancement of optical Kerr nonlinearity if the Rydberg-Rydberg interaction is mapped to photon-photon interaction via EIT~\cite{Fir2016,Mur2016}.

In addition to the giant enhancement, the Kerr nonlinearity in Rydberg gases possesses many other interesting properties. One of them is its nonlocality, originated from the long-range character of the Rydberg-Rydberg interaction~\cite{Sevincli2011,Stanojevic2013,Grankin2015,Bienias2016,Bai2016,
Tebben2019,Bai2019,Sinclair2019}. Based on such nonlocality, Sevincli {\it et al.}~\cite{Sevincli2011} showed that a hexagonal optical pattern can spontaneously form through a modulational instability (MI) of plane-wave probe field in a ladder-shaped three-level Rydberg gas (Rydberg-EIT) with repulsive Rydberg-Rydberg interaction. Recently, it was demonstrated that a structural phase transition of optical patterns from a hexagonal lattice to two types of square lattices may occur in an EIT-based Rydberg gas with a microwave dressing between two Rydberg states~\cite{Shi2020}. These investigations enriched our understanding on the MI and related pattern formation in systems with repulsive (or with both repulsive and attractive) Kerr nonlinearities, which are topics explored in different physical systems by many research groups, from which new pattern formation mechanisms for conservative nonlocal nonlinear systems were found in recent years~\cite{Saito2009,Henkel2010,Cinti2010,Mottl2012,Henkel2012,Hsuch2012,
Hsueh2013,Labeyrie2014,
Cinti2014,Camara2015,Lu2015,Kadau2016,Wachtler2016,Maucher2016,ZhangYC2018,
Firth2017,Hsuch2017,Li2018,Xi2018,Zhang2019}.

In this paper, we consider the formation and manipulation of stationary optical patterns in a cold four-state Rydberg gas with a repulsive Rydberg-Rydberg interaction. The atoms under study have an inverted Y-shaped level configuration, interacting with a control field and a probe field with two orthogonal polarization components and working under condition of double Rydberg-EIT~\cite{Cross1993}. Starting from Maxwell-Bloch (MB) equations and using an approach on atom-atom correlations beyond mean-field approximation, we derive two coupled three-dimensional (3D) nonlocal nonlinear Schr\"odinger (NNLS) equations, in which nonlocal self- and cross-Kerr nonlinearities display specific characters. Based on the MI analysis of plane-wave state of the probe field, we show that the system can undergo a spontaneous symmetry breaking, resulting in the emergence of novel self-organized optical structures.

Through detailed analytical and numerical calculations, we find that the plane-wave state can be transited into many (at least eight) types of self-organized spatial optical lattice patterns. These optical patterns are controlled by the ratio between the cross- and self-Kerr nonlinearities, the nonlocality degree of the  Kerr nonlinearities, and the populations initially prepared in the two atomic ground states. Interestingly, a crossover from spatial mixture to spatial separation (i.e. optical phase separation) for the two polarization components of the probe field may appear when the ratio between the cross- and self-Kerr nonlinearities reaches and beyond a critical value.
We also show that the system can support nonlocal two-component spatial optical solitons and vortices if the parameters of the system are selected suitably to make the signs of the Kerr nonlinearities changed. The rich diversity in types and active controllability in properties of the self-organized optical structures found in this work provide ways for realizing novel optical patterns and solitons and for manipulating their structural phase transitions by exploiting Rydberg gases via EIT.

The remainder of the paper is arranged as follows. In Sec.~\ref{sec2}, we present the physical model of the double Rydberg-EIT under study and derive two coupled 3D NNLS equations describing the evolution of the two polarized components of the probe field beyond mean-field approximation. In Sec.~\ref{sec3}, we consider the MI of a plane-wave state, and investigate the pattern formation, structural phase transition, optical phase separation, and nonlocal optical soliton and vortices in the system. Finally, Sec.~\ref{sec4} gives a summary of the main results obtained in our work.

\section{Model and coupled nonlinear envelope equations}\label{sec2}

\subsection{Physical model}\label{sec2A}

\begin{figure}
\includegraphics[width=1\linewidth]{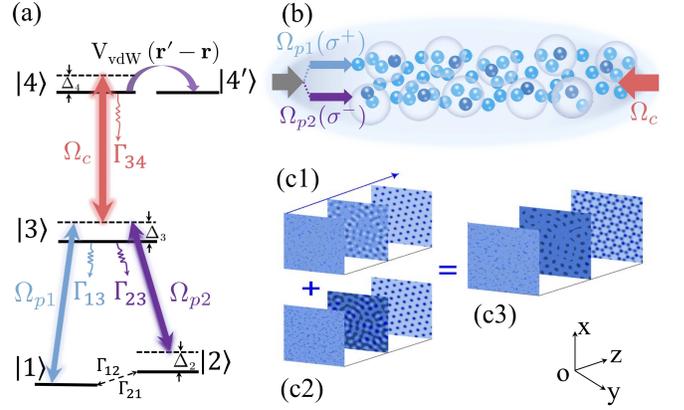}
\caption{\footnotesize
(a)~Atomic level diagram and inverted Y-shaped excitation scheme supporting the double Rydberg-EIT. The $\sigma^+$ ($\sigma^-$) polarization component of the probe field  [half Rabi frequency ${\Omega}_{p1}$ (${\Omega}_{p2}$)] drives the transition $|1\rangle\leftrightarrow|3\rangle$ ($|2\rangle\leftrightarrow|3\rangle$);
the strong control field with half Rabi frequency ${\Omega}_c$ drives the transition $|3\rangle\leftrightarrow|4\rangle$.
$|1\rangle$ and $|2\rangle$: ground states;
$|3\rangle$: intermediate excited state;
$|4\rangle$: highly excited Rydberg state;
$\Delta_{3}$: one-photon detunings;
$\Delta_{2}$ and $\Delta_{4}$: two-photon detunings;
$\Gamma_{13}$, $\Gamma_{23}$, and $\Gamma_{34}$: spontaneous emission decay rates;
$\Gamma_{12}$ and $\Gamma_{21}$: rates of incoherent population transfer between the two ground states; $V_{\mathrm{vdW}}(\mathbf{r}^{\prime}-\mathbf{r}) \equiv \hbar \mathcal{V}(\mathbf{r}^{\prime}-\mathbf{r})$: the van der Waals potential describing the interaction between two Rydberg atoms located respectively at position ${\bf r}$ and ${\bf r}'$.
(b)~Possible experimental geometry, where small solid circles denote
atoms, and large transparent spheres denote Rydberg blockade spheres in which only one atom is excited into Rydberg state.
(c1-c3)~Pattern formation via the modulation instability of a plane-wave probe field, where (c1)  [(c2)] represents the $\sigma^+$ ($\sigma^-$) polarization component and (c3) represents their superposition (total probe field).
}\label{fig1}
\end{figure}

We start to consider an ensemble of lifetime-broadened four-level atoms with an inverted Y-shaped excitation scheme [see Fig.~\ref{fig1}(a)]. Here
%
the $\sigma^+$ ($\sigma^-$) polarization component of a weak probe field ${\bf E}_p$ [with angular frequency $\omega_{p}$, wave vector ${\bf k}_{p+}\, ({\bf k}_{p-})$, and half Rabi frequency ${\Omega}_{p1}$ (${\Omega}_{p2}$)] drives the transition $|1\rangle\leftrightarrow|3\rangle$ ($|2\rangle\leftrightarrow|3\rangle$); the strong, linearly polarized control field with half Rabi frequency ${\Omega}_c$ drives the transition $|3\rangle\leftrightarrow|4\rangle$; $|1\rangle$ and $|2\rangle$ are two ground states, $|3\rangle$ is intermediate excited state, and $|4\rangle$ is a highly excited Rydberg state; $\Delta_{3}$ and $\Delta_{2,4}$ are respectively one- and two-photon detunings; $\Gamma_{13}$, $\Gamma_{23}$, and $\Gamma_{34}$ are spontaneous emission decay rates; $\Gamma_{12}$ and $\Gamma_{21}$ are rates of incoherent population transfer between the two ground states.

The total electric field in the system reads
${\bf E}({\bf r},t)={\bf E}_p+{\bf E}_c=\hat{{\bf e}}_{+}{\mathcal E}_{p+} \exp[i({\bf k}_{+}\cdot{\bf r}-\omega_{p}t)]
+\hat{{\bf e}}_{-}{\mathcal E}_{p-} \exp[i({\bf k}_{-}\cdot{\bf r}-\omega_{p}t)]
+\hat{{\bf e}}_{c}{\mathcal E}_{c} \exp[i({\bf k}_{c}\cdot{\bf r}-\omega_{c}t)]+{\rm c.c.}$,  where
$\hat{\bf e}_\pm=(\hat{{\bf e}}_x\pm i\hat{{\bf e}}_y)/\sqrt{2}$ and  ${\mathcal E}_{p\pm}$  are respectively unit polarization vectors and the envelopes of the $\sigma^\pm$-polarization component of the probe field [$\hat {\bf e}_x$ ($\hat{\bf e}_y$) is the unit vector along  $x$ ($y$) direction], $\hat{\bf e}_c$ and $\mathcal{E}_c$ are respectively the unit polarization vector and amplitude of the control field.
Note that the transition paths $|1\rangle \rightarrow|3\rangle \rightarrow|4\rangle$ and $|2\rangle \rightarrow|3\rangle \rightarrow|4\rangle$ constitute two  ladder-shaped level configurations (each of them displays a Rydberg-EIT), and hence the system supports a double Rydberg-EIT. The interaction between two Rydberg atoms located respectively at ${\bf r}$ and ${\bf r}'$ is described by van der Waals (vdW) potential $V_{\mathrm{vdW}}(\mathbf{r}^{\prime}-\mathbf{r}) \equiv \hbar \mathcal{V}(\mathbf{r}^{\prime}-\mathbf{r})$.  A possible experimental geometry is presented in Fig.~\ref{fig1}(b).

The Hamiltonian of the system is given by $\hat{H}=\mathcal{N}_a\int d^3{r}\hat{\mathcal{H}}_0({\bf r},t)+(\mathcal{N}_a/2)\int d^3{r}\hat{\mathcal{H}}_1({\bf r},t)$. Here $d^3r=dxdydz$,
$\mathcal{N}_a$ is atomic density, $\hat{\mathcal{H}_0}({\bf r},t)$ is the Hamiltonian density describing the atoms and the interaction between the atoms and light fields,
$\hat{\mathcal{H}_1}({\bf r},t)$ is the Hamiltonian density describing the Rydberg-Rydberg interaction.
Under the electric-dipole and rotating-wave approximations, $\hat{\mathcal{H}}_0$ and $\hat{\mathcal{H}}_1$ have the forms
\begin{subequations}\label{Hamiltonian}
\begin{eqnarray}
\hat{{\mathcal H}}_0
&& = -\hbar\sum_{\alpha
=2}^{4}{\Delta_\alpha\hat{S}_{\alpha \alpha}}-\hbar \left(\Omega_{p1}\hat{S}_{13}+\Omega_{p2}\hat{S}_{23}\right.
\nonumber\\
&& \hspace{5mm}+\left.\Omega_c\hat{S}_{34}+{\rm H.c.}\right),\\
\hat{\mathcal{H}}_1
&& =\mathcal{N}_{a} \int{d^3 {r}^{\prime}\hat{S}_{44}({\bf r},t) \hbar \mathcal{V} ({\bf r}^{\prime}-{\bf r}) \hat{S}_{44} ({\bf r}^{\prime},t )},
\end{eqnarray}
\end{subequations}
\noindent where  $\hat{S}_{\alpha\beta}=|\beta\rangle \langle \alpha|\exp\{i[({\bf k}_{\beta}-{\bf k}_{\alpha})\cdot{\bf r}-(\omega_{\beta}-\omega_{\alpha}+\Delta_{\beta}-\Delta_{\alpha})t]\}$
is the atomic transition (for $\alpha\neq \beta$) and population (for $\alpha=\beta$) operators, satisfying the commutation relation
$[\hat{S}_{\alpha\beta}({\bf
r},t),\hat{S}_{\alpha^\prime\beta^\prime}({\bf r}^\prime,t)]=\mathcal{N}_a^{-1}
\delta ({\bf r}-{\bf r}{^\prime})[\delta_{\alpha\beta'}\hat{S}_{\alpha'\beta}({\bf r},t)-\delta_{\alpha'\beta}\hat{S}_{\alpha\beta'}({\bf r'},t)]$  ($\alpha,\beta=1$-$4$).
The  detunings are respectively given by
$\Delta_2=\omega_{p1}-\omega_{p2}-(E_2-E_1)/\hbar=-(E_2-E_1)/\hbar$ (because $\omega_{p1}=\omega_{p2}=\omega_{p}$),
$\Delta_3=\omega_{p}-(\omega_3-\omega_1)$,
and $\Delta_4=\omega_{p}+\omega_c-(\omega_4-\omega_1)$, with $E_{\alpha}=\hbar\omega_\alpha$  the eigen energy of atomic state $|\alpha\rangle$.
The half Rabi frequencies of the probe and control fields are respectively defined by
$\Omega_{p1}=(\hat{{\bf e}}_{p+}\cdot{\bf p}_{31}){\cal E}_{p+}/\hbar$, $\Omega_{p2}=(\hat{\bf e}_{p-}\cdot{\bf p}_{32}){\cal E}_{p-}/\hbar$, and $\Omega_c=(\hat{\bf e}_{c}\cdot{\bf p}_{43}){\cal E}_{c}/\hbar$,
with ${\bf p}_{\alpha\beta}$  the electric-dipole matrix element associated with the transition between $|\alpha \rangle$ and $|\beta\rangle$.
The last term on the right hand side of Eq.~(\ref{Hamiltonian}) is  contributed by the Rydberg-Rydberg interaction with the vdW potential of the form $\hbar\mathcal{V}({\bf r}^{\prime}-{\bf r})= -\hbar C_{6}/|\mathbf{r}'-\mathbf{r}|^{6}$\, ($C_6$ is  dispersion parameter).

The dynamics of the atoms is controlled by the Heisenberg equation of motion for the atomic operators
$\hat{S}_{\alpha\beta}({\bf r},t)$, i.e., $i\hbar\frac{\partial}{\partial t}\hat{S}_{\alpha\beta}({\bf r},t)=[\hat{H}, \hat{S}_{\alpha\beta}({\bf r},t)]$. Taking expectation values on the both sides of this equation, we obtain the optical Bloch equation involving one- and two-body reduced density matrices, which can be cast into the form
\begin{align}\label{Bloch0}
\frac{\partial{\rho}}{\partial t}=-\frac{i}{\hbar} \left[{\hat{ H}_{0}},{\rho}\right]-\Gamma\left[{\rho}\right]+\hat{R}\,[{\rho}_{\rm twobody}].
\end{align}
Here ${\rho} ({\bf r},t)$ is reduced one-body  density matrix (DM) in the single-particle basis $\{|1\rangle, |2\rangle,|3\rangle,|4\rangle\}$, with the matrix elements defined by $\rho_{\alpha\beta}({\bf r},t)\equiv\langle \hat{S}_{\alpha\beta} ({\bf r},t)\rangle$~\cite{note0}; ${\hat{ H}_{0}}=\mathcal{N}_a\int d^3{r}\hat{\mathcal{H}}_0({\bf r},t)$ is the Hamiltonian in the absence of the Rydberg-Rydberg interaction; $\Gamma$ is a $4\times 4$ relaxation matrix describing the spontaneous emission and dephasing. Due to the existence of the Rydberg-Rydberg interaction, two-body reduced DM  [i.e., ${\rho}_{\rm twobody}$  with DM elements $\rho_{\alpha\beta,\mu\nu} ({\bf r}',{\bf r},t)$] is involved in this equation, represented by the last term $\hat{R}\,[{\rho}_{\rm twobody}]$, with $\hat{R}$ a matrix denoting the contribution from the Rydberg-Rydberg interaction. Explicit expressions of Eq.~(\ref{Bloch0}) and a more detailed discussion on it are presented in Appendix~\ref{appA}.

From Eq.~(\ref{Bloch0}) we see that, due to the Rydberg-Rydberg interaction, the evolution of one-body DM elements $\rho_{\alpha\beta}({\bf r},t)$ involves two-body DM elements $\rho_{\alpha\beta,\mu\nu}({\bf r^{\prime},r},t)\equiv\langle\hat{S}_{\alpha\beta}({\bf r^{\prime}},t) \hat{S}_{\mu\nu}({\bf r},t)\rangle$. Thus, Eq.~(\ref{Bloch0}) is not a closed equation. To obtain the solution of the one-body DM elements $\rho_{\alpha\beta}$, we must solve the equations for the two-body DM elements $\rho_{\alpha\beta,\mu\nu}$, which however involve three-body DM elements $\rho_{\alpha\beta,\mu\nu,\delta\sigma}$, and so on. As a result, one obtains a hierarchy of infinite equations for $N$-body DM elements (also called $N$-body correlators; $N=1,2,3,...$) that must be solved simultaneously.
To make the problem solvable, a consistent and effective approach on such a hierarchy of infinite equations involving various orders of many-body correlators is needed. A powerful technique for such approach is the reduced density matrix expansion, by which the hierarchy of the infinite equations is truncated consistently and the problem can be reduced to solve the closed equations for the one- and two-body DM elements~[see Eq.~(\ref{2body})], as shown recently~\cite{Bai2016,Bai2019}.

The probe field is governed by Maxwell equation $\nabla^2 {\bf E}-(1/c^2)\partial^2 {\bf E}/\partial t^2 =[1/(\varepsilon_0 c^2)]\partial^2 {\bf P}/\partial t^2$, with ${\bf P}=
{\cal N}_a  \sum_{j=1}^2{\bf p}_{j3}\rho_{3j}\exp [i({\bf k}_{p\pm}\cdot {\bf r}-\omega_{p} t)]+{\rm c.c.}$ the
electric-polarization intensity. Under paraxial and slowly-varying envelope approximations, the Maxwell equation is reduced to~\cite{Mur2016}
\begin{align}\label{Maxwell}
  &i\left(\frac{\partial}{\partial z}+\frac{1}{c}\frac{\partial}{\partial t}\right)\Omega_{pj}+\frac{c}{2\omega_{p}}\nabla_{\perp}^2\Omega_{pj}+\kappa_{j3}
  \rho_{3j}=0
\end{align}
for the two polarization components ($j=1,2$), where $\nabla_{\perp}^2=\partial^2/\partial x^2+\partial^2/\partial y^2$, $\kappa_{13}\equiv\mathcal{N}_a\omega_{p}|(\hat{\bf e}_{p+ }\cdot{\bf p}_{13})|^2/(2\varepsilon_0c\hbar)$ and $\kappa_{23}\equiv \mathcal{N}_a\omega_{p}|(\hat{\bf e}_{p- }\cdot{\bf p}_{23})|^2/(2\varepsilon_0c\hbar)$ are  coupling constants.
The probe field is assumed to propagate along $z$ direction, i.e., ${\bf
k}_{p\pm}=(0,0,\omega_{p}/c)$. To suppress Doppler effect, the control field is chosen to propagate along  negative $z$ direction, i.e., ${\bf k}_c=(0,0,-\omega_c/c)$.

The model described above is widely applicable as the assumptions made are fulfilled by typical Rydberg gas experiments. For latter considerations where numerical values of system parameters are needed to obtain theoretical results numerically, here we take cold $^{87}$Rb atomic gas as a realistic example.  The assigned atomic levels are~\cite{RbDline}
$|1\rangle=|5 S_{1/2},\,F=1,\, m_F=-1\rangle$,
$|2\rangle=|5 S_{1/2},\,F=1,\, m_F=1\rangle$,
$|3\rangle=|5 P_{3/2},\,F=1,\, m_F=0\rangle$,
and $|4\rangle=|n S_{1/2}\rangle$.
The parameters are
$\Gamma_{13}=\Gamma_{23}=2\pi \times 3.1$\,MHz,
$\Gamma_{34}=2\pi \times 16.7$\,kHz,
$\Gamma_{21}=\Gamma_{12}=2\pi \times  1.0$\,kHz,
$\Delta_2=2\pi\times0.2$~MHz, $\Delta_3=2\pi \times 120 $~MHz,  $\Delta_4=2\pi \times 0.1$~MHz,
$\Omega_{c}=2\pi\times 13\,{\rm MHz}$,
and $\mathcal{N}_a=5.0\times 10^{10}\,{\rm cm^{-3}}$.
For principle quantum number  $n=60$, $C_6=-2\pi\times 140 \,{\rm GHz\, \mu m^6}$ (i.e. the Rydberg-Rydberg interaction is repulsive)~\cite{Pritchard2010,Singer2005}.

\subsection{Enhanced Kerr nonlinearities}\label{sec2B}

Since the probe field is weak, a standard perturbation expansion developed in Refs.~\cite{Bai2016,Bai2019,Zhang2018} can be applied to solve Eq.~(\ref{Bloch0}) by taking $\Omega_{pj}$ as small parameters. Then we can acquire the solutions of Eq.~(\ref{Bloch0}) up to third-order approximation, which are given in the Appendix~\ref{appA}; particularly, the result of the one-body DM elements $\rho_{31}$ and  $\rho_{32}$ exact to third-order approximation can be obtained analytically.
With the results of $\rho_{31}$ and $ \rho_{32}$, we can obtain the optical susceptibility for the $j$th polarization component of the probe field, i.e., $\chi_{j}= \mathcal{N}_a(\hat{\bf e}_{p\pm}\cdot{\bf  p}_{1j})\rho_{3j}/(\varepsilon_0\mathcal{E}_{p\pm})$, given by
{\small \begin{align}\label{chip0}
\chi_{j}=\chi_{j}^{(1)}
             +\Big[\chi_{j,{\rm loc} }^{(3,s)}+\chi_{j,{\rm nloc}}^{(3,s)}\Big]|{\cal E}_{p\pm}|^{2}
             +\Big[\chi_{j,{\rm loc}}^{(3,c)}+\chi_{j,{\rm nloc}}^{(3,c)}\Big]|{\cal E}_{p\mp}|^{2},
\end{align}}\noindent
where $\chi_{j}^{(1)}$,\,$\chi_{j,\rm loc}^{(3,s)}$\,[$\chi_{j,\rm loc}^{(3,c)}$], and $\chi_{j,\rm nloc}^{(3,s)}$\,[$\chi_{j,\rm nloc}^{(3,c)}$] are linear susceptibility, local self-Kerr (cross-Kerr) nonlinear susceptibility, and nonlocal self-Kerr (cross-Kerr) nonlinear susceptibility, respectively. The explicit expressions of these susceptibilities are presented in Appendix~\ref{appB}.

Using the system parameters given in the last subsection, we obtain
$\chi_{j,{\rm loc}}^{(3,s)}\approx \chi_{j,{\rm loc}}^{(3,c)}=(5.81+0.46i)\times 10^{-12}\,{\rm m^2V^{-2}}$, and $\chi_{j,{\rm nloc}}^{(3,s)}\approx \chi_{j,{\rm nloc}}^{(3,c)}=(-6.31+0.21i)\times 10^{-9}\,{\rm m^2V^{-2}}$.
We see that: (i)~The real parts of all the optical Kerr susceptibilities are much larger than their corresponding imaginary parts, which is due to the double EIT effect induced by the control field;
(ii)~The nonlocal Kerr nonlinear susceptibilities are much larger (around three orders of magnitude) than the local Kerr nonlinear susceptibilities. The reason for this is that the strong atom-atom interaction contributed by the Rydberg excitations plays a dominant role over the photon-atom interaction to the nonlinear optical polarization in the system~\cite{Bai2016,Bai2019}.

\subsection{Coupled nonlocal NLS equations}\label{sec2C}

With the results of $\rho_{31}$ and $ \rho_{32}$ obtained by the perturbation expansion exact to the third-order approximation, coupled envelope equations controlling the dynamics of the two polarization components of the probe field can be derived from the Maxwell Eq.~(\ref{Maxwell}), which read
{\small\begin{align}\label{cnls00}
&i\frac{\partial \Omega_{pj} }{\partial z}+\frac{c}{2\omega_{p}}\nabla_{\perp}^{2}\Omega_{pj}+ \sum_{l=1}^2W_{jl}|\Omega_{pl}|^2\Omega_{pj}\notag\\
&+\sum_{l=1}^2\int d^2{r}'\mathcal{N}'_{jl}({\bf r}_{\perp}^{\prime}-{\bf r}_{\perp})|\Omega_{pl}({\bf r}_\perp',z)|^2\Omega_{pj}({\bf r}_\perp,z)=0,
\end{align}}\noindent
for the two polarization components ($j=1,2$), where $d^2r'=dx'dy'$.
Note that, when deriving the above equations, we have assumed that the probe field is a spatial light beam, i.e., its envelopes are stationary~(i.e., $\Omega_{pj}$ is time-independent everywhere). Such assumption is valid for the probe field having a large time duration, so that a continuous-wave (CW) approximation can be applied~\cite{Mur2016,Sevincli2011,Shi2020};
moreover, the spatial width of the envelopes in $z$ direction is assumed to be large so that a local approximation in the $z$ direction for the {\it effective} nonlinear interaction potentials between
photons~\cite{Mur2016,Sevincli2011,Bai2019,Shi2020} (or called spatial response functions) $\mathcal{N}_{jl}({\bf r}^{\prime}-{\bf r})$ can be made, which gives $\int_{-\infty}^{+\infty} \mathcal{N}_{jl}({\bf r}^{\prime} - {\bf r})dz^{\prime}=\mathcal{N}'_{jl}({\bf r}_{\perp}^{\prime}-{\bf r}_{\perp})$. For the detailed derivation of the coupled 3D NNLS equations (\ref{cnls00}), see the Appendix~\ref{appB}.

Shown in Fig.~\ref{fig2}
\begin{figure}
\centering
\includegraphics[width=0.35\textwidth]{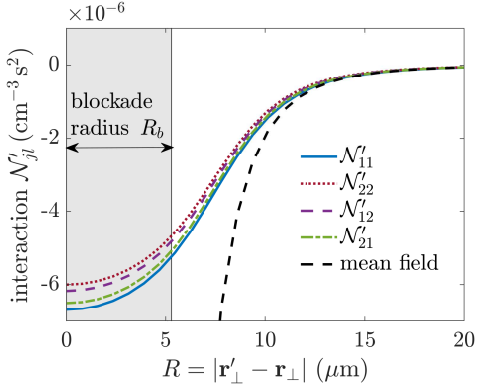}
\caption{\footnotesize Effective nonlinear interaction potentials between photons $\mathcal{N}'_{jl}$ as functions of the separation $R=|{\bf r}'_\perp-{\bf r}_\perp|$. Curves illustrated respectively by solid blue, dotted red, dashed purple, and dotted-dashed green lines are for $\mathcal{N}'_{11}$, $\mathcal{N}'_{22}$, $\mathcal{N}'_{12}$,  and $\mathcal{N}'_{21}$, which are approach finite values when $R \rightarrow 0$. For comparison, the result obtained by using mean-field approximation is also shown (dashed black line), which diverges rapidly when $R$ becomes small.
The shadow region on the left hand side is the one for small interatomic distance, where the effective interaction tends to be finite and saturated due to the atomic correlations.
}
\label{fig2}
\end{figure}
are $\mathcal{N}'_{jl}$ as functions of the interatomic separation $R=|{\bf r}'_\perp-{\bf r}_\perp|$, calculated by using the parameters given in subsection \ref{sec2A}. Curves plotted  by solid blue, dotted red, dashed purple, and dotted-dashed green lines are for $\mathcal{N}'_{11}$, $\mathcal{N}'_{22}$, $\mathcal{N}'_{12}$, and $\mathcal{N}'_{21}$, respectively. We see that all the four nonlinear response functions approach finite values for small $R$ (indicated by the shadow region on the left hand side of the figure). This is contributed from the Rydberg-Rydberg interaction that induces strong correlations between the atoms and makes the effective interaction potentials between photons be saturated for small $R$. In the figure, the region of the Rydberg blockade (with radius $R_b$) for the atom-atom interaction is indicated by the shadow one. For comparison, the result obtained by using a mean-field approximation is also shown in the figure (see the dashed black line), which is divergent for small $R$ due to the improper neglect of the atomic correlations in such calculation.

Since the local Kerr nonlinear susceptibilities are much smaller than the nonlocal ones, the third term on the left side of Eqs.~(\ref{cnls00}) plays no significant role and can be safely neglected. For the convenience of the following physical discussions and numerical simulations, we recast Eqs.~(\ref{cnls00}) into the dimensionless form
\begin{align}\label{cnls}
&i\frac{\partial u_j}{\partial s}+\tilde{\nabla}_\bot^2u_j\notag\\
&+\sum_{l=1,2}\int d^2\zeta'\big[\Re_{jl}(\vec{\zeta}'-\vec{\zeta})|u_l(\vec{\zeta}',s)
|^2\big]u_j(\vec{\zeta},s)=0,
\end{align}
where $u_j=\Omega_{pj}/U_0$, $s=z/(2L_{\rm diff})$, $\tilde{\nabla}_\bot^2=\partial^2/\partial\xi^2+\partial^2/\partial\eta^2$, $\vec{\zeta}=(\xi,\eta)=(x,y)/R_0$,  $d^2\zeta'=d\xi'd\eta'$, and $\Re_{jl}(\vec{\zeta}'-\vec{\zeta})=2L_{\rm diff}R_0^2U_0^2\mathcal{N}'_{jl}[(\vec{\zeta}'-\vec{\zeta})R_0]$~\cite{note99},
with $U_0$, $L_{\rm diff}\equiv\omega_pR_0^2/c$, and $R_0$ typical half Rabi frequency, diffraction length, and transverse size (e.g. the lattice separation of optical patterns) of the probe field, respectively~\cite{note100}.  Eqs.~(\ref{cnls}) can be cast into the form
\begin{align}\label{cnls3}
&i\frac{\partial v_j}{\partial s}+\tilde{\nabla}_\bot^2v_j\notag\\
&+\sum_{l=1,2}\int d^2\zeta'\big[I_l\Re_{jl}(\vec{\zeta}'-\vec{\zeta})|v_l(\vec{\zeta}',s)|^2\big]v_j(\vec{\zeta},s)=0,
\end{align}
where $I_j=\int |u_j(\vec{\zeta})|^2d^2\zeta$ is the power of $j$th polarization component; $v_j(\vec{\zeta})= u_j(\vec{\zeta})/I_j^{1/2}$, which satisfies the normalization condition $\int |v_j(\vec{\zeta})|^2d^2\zeta=1$. The property of the reduced response functions $\Re_{jl}$ depend significantly on the {\it nonlocality degree} of the Kerr nonlinearity, defined by
\begin{equation}\label{NLD}
\sigma\equiv {R_b}/{R_0}.
\end{equation}
Here $R_b$ is the blockade radius of Rydberg blockade sphere, given by
$R_b=|C_{6}/\delta_{\rm EIT}|^{1/6}$~\cite{Fir2016,Mur2016}, with
$\delta_{\rm EIT}$ the width of EIT transparency window. One has
$\delta_{\rm EIT}=|\Omega_c|^2/\gamma_{31}$ for
$\Delta_3=0$, and $\delta_{\rm EIT}=|\Omega_c|^2/\Delta_3$
for $\Delta_3\gg\gamma_{31}$. With the system parameters used here,
we have $R_b\approx 5.40 ~\mu{\rm m}$. The value of $\sigma$ can be varied
by changing $R_0$, used in the following calculations.

\section{Modulation instability, pattern formation, phase separation,  solitons and vortices}\label{sec3}

\subsection{Modulation instability for the two polarization components of the probe field}\label{sec3A}

We now consider the MI and related pattern formation that may occur in the system based on the coupled NNLS equations obtained above.
MI is a nonlinear instability of constant-amplitude continuous waves under the influence of small perturbations, occurring in a variety of contexts where Kerr nonlinearities are attractive and local~\cite{Zakharov2009,Biondini2016}; it can also arise in systems with repulsive but nonlocal Kerr nonlinearity when the perturbations have both long~\cite{Krolikowski2001,Krolikowski2004} and short~\cite{Sevincli2011,Maucher2016} wavelengths.
%
\begin{figure*}[htpb]
\centering
\includegraphics[width=0.97\textwidth]{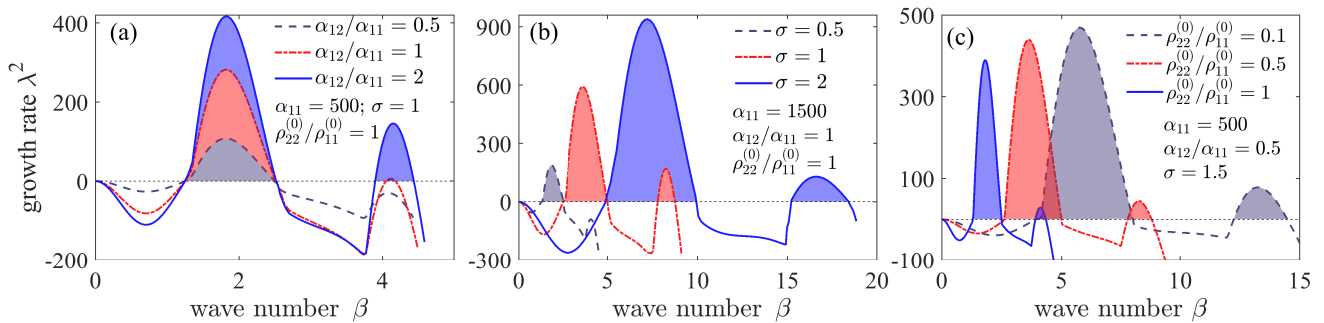}
\caption{\footnotesize Growth rate of the MI for the two-component plane-wave probe field as a function of the dimensionless wave number $\beta=\sqrt{\beta_1^2+\beta_2^2}$ [$\beta_1\equiv R_0 k_x$, $\beta_2\equiv R_0 k_y$] for different nonlinearity parameters $\alpha_{jl}$ [defined in Eq.~(\ref{effectnon}), with $\alpha_{11}=\alpha_{22}$, $\alpha_{12}=\alpha_{21}$] and the nonlocality degree $\sigma = R_b/R_0$ of the Kerr nonlinearities.  Here $k_x$, $k_y$ are respectively wave numbers in $x$ and $y$ directions; $R_0$ and  $R_b$  are respectively the typical transverse beam radius of the probe field and the radius of Rydberg blockade sphere.
(a)~$\lambda^2$  as a function of $\beta$ for the cross-Kerr nonlinearity parameter $\alpha_{12}=0.5\,\alpha_{11}$~(dashed gray line), $\alpha_{12}=\alpha_{11}$~(dotted red line), and $\alpha_{12}=2\,\alpha_{11}$~(solid  blue line), respectively. Here $\sigma =1$, $\alpha_{11}=500$, and $\rho_{22}^{(0)}/\rho_{11}^{(0)}=1$  ($\rho_{11}^{(0)}$ and $\rho_{22}^{(0)}$ are the initial populations
at the two atomic ground states) are assumed.
(b)~$\lambda^2$  as a function of  $\beta$
for the nonlocality degree of the Kerr nonlinearities
$\sigma = 0.5$ (dashed gray line), $\sigma =1$ (dotted red line), and $\sigma=2$~(solid  blue line), respectively. Here $\alpha_{11}=1500$, $\alpha_{12}=\alpha_{11}$, and $\rho_{22}^{(0)}/\rho_{11}^{(0)}=1$ are assumed.
(c)~$\lambda^2$  as a function of  $\beta$ for $\rho_{22}^{(0)}/\rho_{11}^{(0)}=0.1$ (dashed gray line), $\rho_{22}^{(0)}/\rho_{11}^{(0)} =0.5$ (dotted red line), and $\rho_{22}^{(0)}/\rho_{11}^{(0)}=1$~(solid blue line), respectively. Here $\alpha_{11}=500$, $\alpha_{12}=0.5\alpha_{11}$, and $\sigma=1.5$ are assumed.
Colorful regions in all the panels are ones for Re$(\lambda) > 0$, where MI occurs and hence the two-component plane-wave probe field is unstable~\cite{note1}.
}\label{fig3}
\end{figure*}

To explore the MI in the present system, we consider the plane-wave solution of the coupled NNLS Eqs.~(\ref{cnls3}), i.e.,
$v_{{\rm pw},j}(\vec{\zeta},s)=v_{j0}\exp(i\mu_js) $
with $\mu_j=\iint d^2\zeta\big[I_jv_{10}^2\Re_{jj}(\vec{\zeta})+I_{3-j} v_{20}^2\Re_{3-j,j}(\vec{\zeta})]$ ($j=1,2$). Since any perturbation can be expanded as a superposition of many Fourier modes, the modulation of the plane-wave solution by the perturbation can be cast into the  form
\begin{align}\label{PTS}
\tilde{v}_j(\vec{\zeta},s)=
&\left[v_{j0}+a_{1j}e^{i\vec{\beta}\cdot \vec{\zeta}+\lambda
s}+a_{2j}^*e^{-i\vec{\beta}\cdot \vec{\zeta}+\lambda^* s}\right]e^{i\mu_js},
\end{align}
where $a_{1j}$ and $a_{2j}$ are small complex amplitudes of the perturbation, $\vec{\beta}= (\beta_1,\beta_2)$\, ($\beta_1\equiv R_0 k_x$, $\beta_2\equiv R_0 k_y$; $k_x$, $k_y$ are wave numbers in $x$ and $y$ directions, respectively) is dimensionless 2D wave vector, and $\lambda$ is the growth rate of the perturbation.

Substituting the perturbation solution (\ref{PTS}) into  Eqs.~(\ref{cnls3}) and keeping only linear terms of $a_{1j}$ and $a_{2j}$, we can obtain $\lambda=\lambda_{\pm}$. The result is given by
\begin{align}\label{MI}
\lambda_\pm ^2&=\beta^2A_{+}-\beta^4 \pm\beta^2\sqrt{A_{-}^2+4I_1I_2\tilde\Re_{12}(\vec\beta)\tilde\Re_{21}(\vec\beta)},
\end{align}
where $A_\pm= I_1\tilde\Re_{11}(\vec\beta)\pm I_2\tilde\Re_{22}(\vec\beta)$, ${\beta}=\sqrt{\beta_1^2+\beta_2^2}$, and $\tilde{\Re}_{jl}(\vec{\beta})$ is the response function in momentum space [i.e., the Fourier transformation of $\Re_{jl}(\vec{\zeta})$].
We find the growth rate $\lambda$ has four branches, i.e., $\lambda_{+(1,2)}$ and $\lambda_{-(1,2)}$, in which only the positive branch can result in MI.  The property of $\lambda$ depends on the power of the input probe fields $I_j$ and the shape of the response function $\tilde{\Re}_{jl}$. For convenience, we define the {\it nonlinearity parameters}
\begin{align}\label{effectnon}
\alpha_{jl}= {I}_l\int \Re_{jl}(\vec{\zeta})d^2\zeta.
\end{align}
Due to the symmetry of the inverted Y-type level structure considered here [Fig.~\ref{fig1}(a)\,], if $I_1=I_2$ one has approximately  $\alpha_{11}=\alpha_{22}$ and $\alpha_{12}=\alpha_{21}$.
Other cases for  $I_1\neq I_2$ may give very rich behaviors of the pattern formation for the two polarized components of the probe field.  In this work, for simplicity, we discuss only the particular case $I_1=I_2$.

The MI can be controlled by manipulating the physical parameters of the system. Fig.~\ref{fig3}
shows $\lambda^2$ as a function of the dimensionless wave number $\beta$ for different nonlinearity parameters $\alpha_{jl}$, the nonlocality degree $\sigma$, and the {\it ratio of initial populations} prepared in two atomic ground states, i.e. $\rho_{22}^{(0)}/\rho_{11}^{(0)}$~\cite{note1}.
Plotted in Fig.~\ref{fig3}(a) is the curve of $\lambda^2$  as a function of $\beta$ for the cross-Kerr nonlinearity parameter $\alpha_{12}=0.5\,\alpha_{11}$~(dashed gray line), $\alpha_{12}=\alpha_{11}$~(dotted red line), and $\alpha_{12}=2\,\alpha_{11}$~(solid blue line), respectively (here $\sigma =1$, $\alpha_{11}=500$, and $\rho_{22}^{(0)}/\rho_{11}^{(0)}=1$  are assumed). We see that the MI may happen in different domains of the wave number $\beta$. The reason for the appearance of the second MI domain stems from the cross-Kerr nonlinearity of the system.

The nonlocality degree of the Kerr nonlinearities can also be used to control the MI. Fig.~\ref{fig3}(b) shows $\lambda^2$  as a function of wave number $\beta$ for the nonlocality degree
$\sigma = 0.5$ (dashed gray line), $\sigma =1$ (dotted red line), and $\sigma=2$~(solid  blue line), respectively (here $\alpha_{11}=1500$, $\alpha_{12}=\alpha_{11}$, and $\rho_{22}^{(0)}/\rho_{11}^{(0)}=1$  are assumed). In this case the MI domain is significantly modified when $\sigma$ is varied, which is quite different from those modifications induced by the Kerr nonlinearities illustrated in Fig.~\ref{fig3}(a); furthermore, the critical wave number for the appearance of the MI are also enlarged greatly.

The populations initially prepared in the two atomic ground states $|1\rangle$ and $|2\rangle$  (i.e. $\rho_{11}^{(0)}$, $\rho_{22}^{(0)}$) also play an important role for the appearance of the MI.
Plotted in Fig.~\ref{fig3}(c) is $\lambda^2$ as a function of $\beta$ for the ratio of the populations $\rho_{22}^{(0)}/\rho_{11}^{(0)}=0.1$ (dashed gray line), $\rho_{22}^{(0)}/\rho_{11}^{(0)} =0.5$ (dotted red line), and $\rho_{22}^{(0)}/\rho_{11}^{(0)}=1$~(solid blue line), respectively (here $\alpha_{11}=500$, $\alpha_{12}=0.5\alpha_{11}$, and $\sigma=1.5$ are assumed). In this situation the MI domain is dramatically changed when $\rho_{22}^{(0)}/\rho_{11}^{(0)}$ is varied; moreover, the critical wave number for the appearance of the MI decreases with the increase of $\rho_{22}^{(0)}/\rho_{11}^{(0)}$.

From these results we see that the MI depends not only on the nonlinearity parameters ($\alpha_{jl}$), but also on the nonlocality degree of Kerr nonlinearities ($\sigma$) as well as the population initially prepared in the two atomic ground states ($\rho_{22}^{(0)}/\rho_{11}^{(0)}$). This provides many ways to manipulate the MI and thereby the optical patterns in the system, discussed in the next subsections.

\subsection{Pattern formations controlled by the ratio between the cross- and self-Kerr nonlinearities}\label{sec3B}

We now explore the outcome of the MI discussed above.
To demonstrate the appearance of optical patterns, we seek the stationary-state solutions of the system, for which the total energy
\begin{align}\label{energy}
E &= \sum_{j=1}^2\int |\tilde{\nabla}_\perp v_j|^2d^2\zeta+\sum_{j,l=1}^2E_{jl}
\end{align}
is minimal. Such solutions of Eqs.~(\ref{cnls3}) are sought in the form $v_{j}=v_{j0}\exp(iE_0 s)$ (with $E_0$ the ground-state energy), which can be obtained numerically by using an imaginary propagation together with split-step Fourier method~\cite{YangJK2010} (for details, see Appendix \ref{appD}). In the above expression, the first term is kinetic energy and the second term
[with $E_{jl}\equiv\frac{1}{2}\iint  d^2\zeta d^2\zeta'  \Re_{jl}(\vec{\zeta}'-\vec{\zeta})|v_j(\vec{\zeta}',s)|^2|v_l(\vec{\zeta},s)|^2$]
is interaction energy. The initial condition used in the numerical simulation is a plane wave, perturbed by a random noise.

%
\begin{figure*}[t]
\centering
\includegraphics[width=0.98\textwidth]{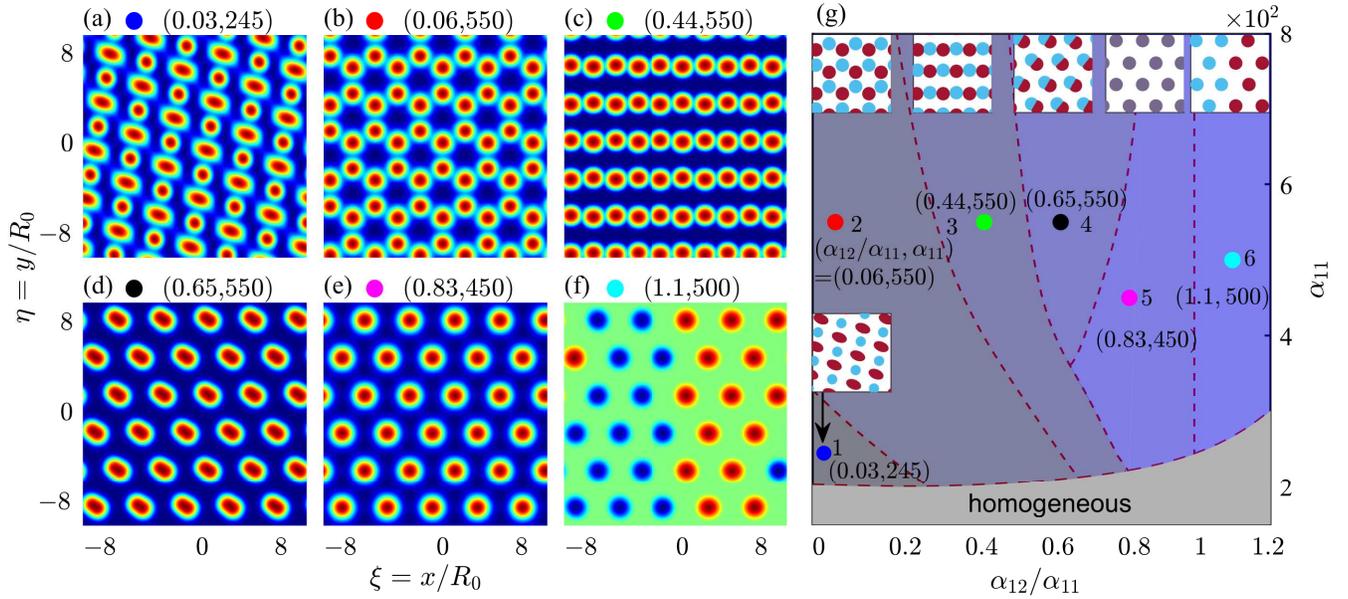}
\caption{\footnotesize
Pattern formation of the two-component probe field controlled by the nonlinearity parameters $\alpha_{11}$ and $\alpha_{12}$, by taking the dimensionless normalized probe-field intensity $|v|^2\equiv|v_1|^2+|v_2|^2$
as a function of the dimensionless coordinates $\xi = x/R_0$ and $\eta = y/R_0$ (with $\sigma=1$ and $\rho_{22}^{(0)}/\rho_{11}^{(0)}=1$) for different nonlinearity parameters:
(a)~($\alpha_{12}/\alpha_{11}, \alpha_{11}) = (0.03, 245$);
(b)~($\alpha_{12}/\alpha_{11}, \alpha_{11}) = (0.06, 550$);
(c)~($\alpha_{12}/\alpha_{11}, \alpha_{11}) = (0.44, 550$);
(d)~($\alpha_{12}/\alpha_{11}, \alpha_{11}) = (0.65, 550$);
(e)~($\alpha_{12}/\alpha_{11}, \alpha_{11}) = (0.83, 450$);
(f)~($\alpha_{12}/\alpha_{11}, \alpha_{11}) = (1.1, 550$).
In the figure, the region with brighter (darker) color means that the light intensity in that region is higher (lower).
(g)~Phase diagram for the optical pattern formation, in which different domains (phases) are obtained by using different values of ($\alpha_{12}/\alpha_{11}$,$\alpha_{11}$) corresponding respectively to those used in (a)-(f). The inserted schematics in each domain stands for the optical pattern obtained with the indicated value of ($\alpha_{12}/\alpha_{11}$,$\alpha_{11}$), where the distributions of two probe-field components are illustrated by the solid blue circles (denoting $|v_1|^2$) and solid red circles (denoting $|v_2|^2$), respectively. Large solid circles with different colors in six domains (with numbers 1,\,2,\,3,\,4,\,5,\,6 indicated respectively) of the phase diagram are used to guide the look for the correspondence between different phases and different optical patterns given in (a)-(f), respectively. Note that a spatial phase separation occurs in the case shown in (f), where the light-intensity distributions of the two polarization components are separated in space.
}\label{fig4}
\end{figure*}

Due to the spontaneous symmetry breaking induced by the MI, the plane-wave state of the system is transferred into new states, manifested by the emergence of structured optical patterns. Fig.~\ref{fig4} shows the optical patterns [Fig.~\ref{fig4}~(a)-(f)] and the related phase diagram [Fig.~\ref{fig4}(g), where ``homogeneous'' means the plane-wave state] of the two-component probe field controlled by the nonlinearity parameters $\alpha_{11}$ and $\alpha_{12}$ [$\alpha_{jl}$ is defined in Eq.~(\ref{effectnon})], by taking the dimensionless, normalized field intensity $|v|^2\equiv|v_1|^2+|v_2|^2$~\cite{note2} as a function of the dimensionless coordinates $\xi=x/R_0$ and $\eta = y/R_0$ (with $\sigma=1$ and $\rho_{22}^{(0)}/\rho_{11}^{(0)}=1$). In Fig.~\ref{fig4}~(a)-(f), the region with brighter (darker) color mean that the light intensity in that region is higher (lower). For convenience,
in each domain of Fig.~\ref{fig4}(g) an insertion is provided to distinguish the distributions $|v_1|^2$ and $|v_2|^2$, denoted by solid blue and red circles, respectively.

Shown in Fig.~\ref{fig4}(a) is the optical pattern for the case of ($\alpha_{12}/\alpha_{11},\alpha_{11})=(0.03, 245)$, where the circular and elliptical spots are for the $v_1$ (i.e., $\sigma^+$) and $v_2$ (i.e., $\sigma^-$) polarization components, respectively. We see that both the $v_1$ and $v_2$ components form hexagonal lattices, which locate at different positions. An obvious feature is that the spots of the $v_2$-component are elliptical but those of the $v_1$-component are circular. The reason for the appearance of such a phenomenon is that the $v_1$-component gives a hexagonal lattice structure with circular bright spots (where light intensity is maximal) in a dark background, the $v_2$-component gives a hexagonal lattice structure but with circular dark spots~(where light intensity is minimal) in a bright background. As a result, the intensity superposition of the two polarized components (i.e. $|v_1|^2+|v_2|^2$) gives an optical pattern in which the circular and elliptical spots appear alternatively in space. For details, see Appendix \ref{app3}. This case corresponds to the domain 1 of the phase diagram given in Fig.~\ref{fig4}(g).

Figure~\ref{fig4}(b) shows the result for $(\alpha_{12}/\alpha_{11},\alpha_{11})=(0.06,550)$. The optical patterns for both the $v_1$ and $v_2$ components obtained here are also hexagonal lattices, but their superposition $|v_1|^2+|v_2|^2$ forms a honeycomb lattice, i.e. one circular spot of one component is surrounded by three circular spots of the other component. This situation is relevant to the domain 2 in the phase diagram shown by Fig.~\ref{fig4}(g). As an example, we also show the formation process of the hexagonal lattices for $v_1$ and $v_2$ components and the honeycomb lattice of their superposition  $|v_1|^2+|v_2|^2$ via the MI starting from an inhomogeneous plane-wave state, see Fig.~\ref{fig1}(c1)-(c3) where (c1)  [(c2)] represents the $\sigma^+$\, ($\sigma^-$) polarization component and (c3) represents their superposition (i.e. the total probe field) for different propagation distance $z$.

The ratio of nonlinearity $\alpha_{12}/\alpha_{11}$ plays an important role for forming different optical structures. To test this idea, we increase $\alpha_{12}$ to make the nonlinearity parameters adjusted to be $(\alpha_{12}/\alpha_{11},\alpha_{11})=(0.44,550)$. In this case, a different optical structure emerges in the system, as illustrated in Fig.~\ref{fig4}(c), where two hexagonal lattices (each consists of circular spots) are staggered each other and they form a square lattice pattern for their superposition $|v_1|^2+|v_2|^2$. This case corresponds to the domain 3 of the phase diagram given in Fig.~\ref{fig4}(g).

Shown in  Fig.~\ref{fig4}(d) and Fig.~\ref{fig4}(e) are for
$(\alpha_{12}/\alpha_{11},\alpha_{11})$ equaling to $(0.65, 550)$ and $(0.83, 450)$, respectively. Both cases still give hexagonal lattice patterns; the lattice spots in Fig.~\ref{fig4}(d) are elliptical, while the lattice spots in Fig.~\ref{fig4}(e) are circular. The reason is that in Fig.~\ref{fig4}(d) the lattice-point positions of the $v_1$-component have a small separation from those of the $v_2$-component, while in Fig.~\ref{fig4}(e) the lattice-point positions of the $v_1$-component coincide with those of the $v_2$-component. This point can be clearly seen from the phase diagram of these two cases [i.e., the domains 4 and 5 of Fig.~\ref{fig4}(g)], where  the light-intensity distributions of $|v_1|^2$ and $|v_2|^2$ have been given by the inserted schematics.

Interestingly, when the cross-Kerr nonlinearity is increased further, i.e.
$(\alpha_{12}/\alpha_{11},\alpha_{11})=(1.1, 500)$, an optical phase separation occurs in the system, as shown by Fig.~\ref{fig4}(f), where the light-intensity distributions of $|v_1|^2$ (displayed by solid blue circles) and $|v_2|^2$ (displayed by solid red circles) are separated in space, which  is relevant to the domain 6 in Fig.~\ref{fig4}(g). Such phase separation phenomenon is very similar to that occurring in a mixture of binary fluids~\cite{Glotzer1994,Timmermans1998,Ao1998,Pooley2005,Sabbatini2011,Wen2012,
Shevtsova2015,Cates2018,Ota2019,Kumar2019}. We shall discuss this topic in more detail in Sec.~\ref{sec3E}.

\subsection{Pattern formations controlled by the nonlocality degree of the Kerr nonlinearities}\label{sec3C}
The nonlocality degree of Kerr nonlinearities can also be used to manipulate the MI [see Fig.~\ref{fig3}(b)], trigger symmetry breaking and generate different optical patterns in the system.
Similar to the last subsection, one can find the stationary-state distributions of the probe field [for which the total energy (\ref{energy}) is minimal] through a numerical simulation on Eq.~(\ref{cnls3}) for different values of the nonlocality degree of Kerr nonlinearities.

\begin{figure*}
  \centering
  \includegraphics[width=0.98\textwidth]{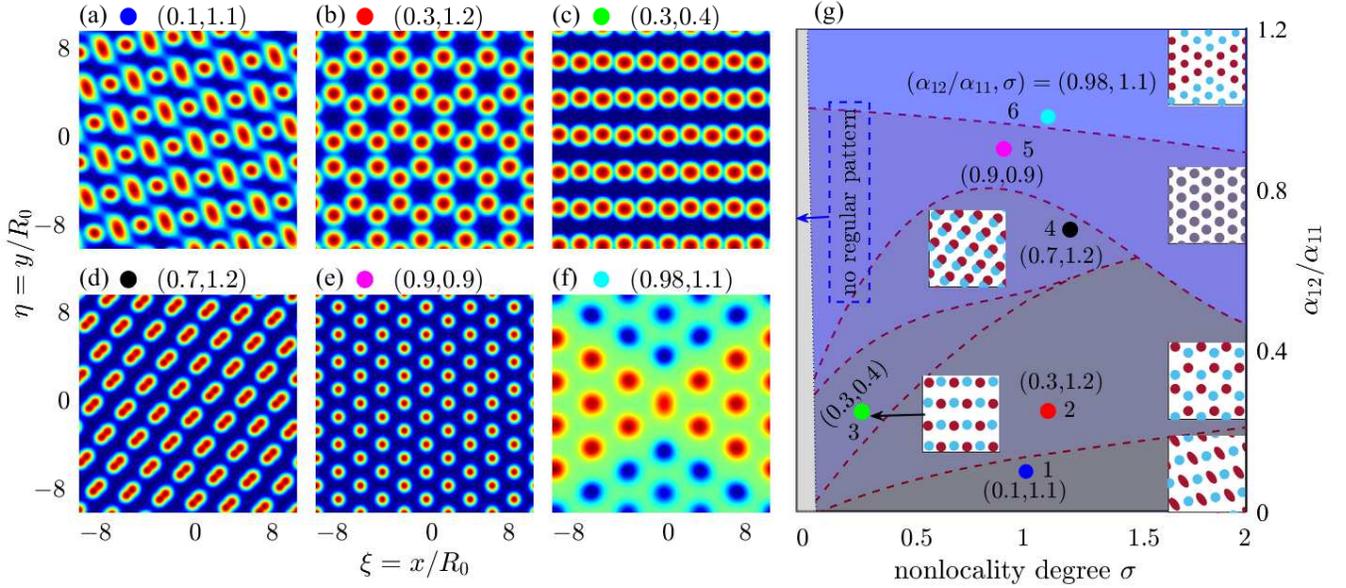}
\caption{\footnotesize Pattern formation of the two-component probe field controlled by the ratio of nonlinearity $\alpha_{12}/\alpha_{11}$ and the nonlocality degree $\sigma$, by taking the dimensionless normalized probe-field intensity $|v|^2=|v_1|^2+|v_2|^2$ as a function of the dimensionless coordinates $\xi = x/R_0$ and $\eta = y/R_0$ (with  $\alpha_{11}=550$  and $\rho_{22}^{(0)}/\rho_{11}^{(0)}=1$) for different $\alpha_{12}/\alpha_{11}$  and $\sigma$:
(a) $(\alpha_{12}/\alpha_{11}, \sigma)= (0.1, 1.1)$;
(b) $(\alpha_{12}/\alpha_{11}, \sigma)= (0.3, 1.2)$;
(c) $(\alpha_{12}/\alpha_{11}, \sigma)= (0.3, 0.4)$;
(d) $(\alpha_{12}/\alpha_{11}, \sigma)= (0.7, 1.2)$;
(e) $(\alpha_{12}/\alpha_{11}, \sigma)= (0.9, 0.9)$;
(f) $(\alpha_{12}/\alpha_{11}, \sigma)= (0.98, 1.1)$.
(g)~Phase diagram for the optical pattern formation, in which different domains (phases) are obtained by using different values of ($\alpha_{12}/\alpha_{11}$,\,$\sigma$) corresponding respectively to those used in (a)-(f).  The inserted schematics in each domain stands for the optical pattern obtained with the indicated value of ($\alpha_{12}/\alpha_{11}$,\,$\sigma$), where the distributions of two probe-field components are illustrated by the solid blue circles (denoting $|v_1|^2$) and solid red circles (denoting $|v_2|^2$), respectively. Large solid circles with different colors in six domains (with numbers 1,\,2,\,3,\,4,\,5,\,6 indicated respectively) of the phase diagram are used to guide to look for the correspondence between different phases and different optical patterns given in (a)-(f), respectively. A spatial phase separation occurs in the case shown in (f), where the distributions of the two polarization components are separated in space (the solid blue circles are for $|v_1|^2$ and solid red circles are for $|v_2|^2$). Note that no regular pattern occurs in the thin domain (gray color) on the leftmost side of (g) where $\sigma$ is very small~(i.e., about $\sigma<0.05$).}
\label{fig5}
\end{figure*}
Figure~\ref{fig5} shows the optical patterns [Fig.~\ref{fig5}~(a)-(f)] and the related phase diagram [Fig.~\ref{fig5}(g)] of the two-component probe field controlled by the ratio of nonlinearity $\alpha_{12}/\alpha_{11}$ and the nonlocality degree $\sigma$, by taking $|v_1|^2+|v_2|^2$ as a function of $\xi=x/R_0$ and $\eta = y/R_0$  (with $\alpha_{11}=550$ and $\rho_{22}^{(0)}/\rho_{11}^{(0)}=1$). Similar to Fig.~\ref{fig4}(g),
in each domain of Fig.~\ref{fig5}(g), an insertion is given for distinguishing the distributions  $|v_1|^2$ (denoted by solid blue circles) and $|v_2|^2$ (denoted by  solid red circles).

Plotted in Fig.~\ref{fig5}(a) is the intensity distribution $|v_1|^2+|v_2|^2$  of the optical pattern for the case of ($\alpha_{12}/\alpha_{11},\sigma)=(0.1, 1.1)$. We see that both the $v_1$ and $v_2$ components form hexagonal lattices, locating at different positions, but the circular and elliptical spots are for the $v_1$ (i.e., $\sigma^+$) and $v_2$ (i.e., $\sigma^-$) polarization components, respectively, which is similar with that of Fig.~\ref{fig4}(a) [see the explanation given in the Appendix \ref{app3}].  This case corresponds to the domain 1 of the phase diagram given in Fig.~\ref{fig5}(g).

Shown in Fig.~\ref{fig5}(b) is the result for  ($\alpha_{12}/\alpha_{11},\sigma)=(0.3, 1.2)$, which is similar with that of in Fig.~\ref{fig4}(b). The optical patterns for both the $v_1$ and $v_2$ components obtained are also hexagonal lattices, but their superposition $|v_1|^2+|v_2|^2$ forms a honeycomb lattice. This situation is relevant to the domain 2 in the phase diagram shown by Fig.~\ref{fig5}(g).

Compared to Fig.~\ref{fig5}(b), a different optical structure emerges for ($\alpha_{12}/\alpha_{11},\sigma)=(0.3, 0.4)$, as shown in Fig.~\ref{fig5}(c).  Here two hexagonal lattices~(each consists of circular spots) are staggered and they form a square lattice pattern for their superposition $|v_1|^2+|v_2|^2$. This case corresponds to the domain 3 of the phase diagram given in Fig.~\ref{fig5}(g). However, if we further reduce the nonlocality degree $\sigma$ to be smaller than 0.05, no regular optical pattern appears, corresponding to the gray domain on the leftmost side of the phase diagram [Fig.~\ref{fig5}(g)]. The reason for the disappearance of the pattern is due to the fact that the interaction between photons [characterized by the response function $\Re_{jl}(\vec{\zeta}'-\vec{\zeta})$ in Eqs.~(\ref{cnls3})]
becomes a short-ranged, which does not support MI for the repulsive Rydberg-Rydberg interaction and hence the formation of optical patterns is not possible.

Fig.~\ref{fig5}(d) and Fig.~\ref{fig5}(e) show the results for the case ($\alpha_{12}/\alpha_{11},\sigma)=(0.7, 1.2)$ and ($\alpha_{12}/\alpha_{11},\sigma)=(0.9, 0.9)$, respectively. We see that the both cases give hexagonal patterns with circular spots for the $v_1$ and $v_2$ components; the lattice spots for their superposition $|v_1|^2+|v_2|^2$ are elliptical in Fig.~\ref{fig5}(d), but circular in Fig.~\ref{fig5}(e). The reason for the elliptical lattice pattern displayed in Fig.~\ref{fig5}(d) is that the lattice-point positions of the $v_1$-component have a small separation from those of the $v_2$-component; in the case of Fig.~\ref{fig5}(e), however,  the lattice-point positions of the $v_1$-component coincide with those of the $v_2$-component. One can see this point clearly from  the domains 4 and 5 of  Fig.~\ref{fig5}(g), where  the light-intensity distributions of $|v_1|^2$~(solid blue circles) and $|v_2|^2$~(solid red circles) have been given by the inserted schematics.

Illustrated in Fig.~\ref{fig5}(f) is for the case ($\alpha_{12}/\alpha_{11},\sigma)=(0.98, 1.1)$. In this situation, an optical phase separation happens, by which the light-intensity distributions of $|v_1|^2$ (displayed by solid blue circles) and $|v_2|^2$ (displayed by solid red circles) are separated in space. This situation corresponds to the domain 6 in the phase diagram given in Fig.~\ref{fig5}(g). We see from the top part of the phase diagram that the domain of the phase separation is enlarged when the nonlocality degree $\sigma$ increases. This finding tells us that, in addition to the cross-Kerr nonlinearity, the nonlocality degree of the Kerr nonlinearities can also be employed to control the optical phase separation in the system.

\subsection{Pattern formations controlled by the initial populations in the two atomic ground states}\label{sec3D}

We now turn to explore what will happen when the preparation of the initial populations  in the two atomic ground states are changed. The initial populations in the ground states $|1\rangle$ and $|2\rangle$ [see Fig.~\ref{fig1}(a)] are respectively given by $\rho_{11}^{(0)}$ and $\rho_{22}^{(0)}$; except for the normalized condition
(i.e. $\rho_{11}^{(0)}+\rho_{22}^{(0)}=1$), they are quite arbitrary.
In the above discussions, we have taken $\rho_{22}^{(0)}/\rho_{11}^{(0)}=1$ (i.e., $\rho_{11}^{(0)}=\rho_{22}^{(0)}=0.5$) for simplicity.
Because different initial populations can be prepared experimentally (e.g. by optical pumping or applying a microwave field to couple $|1\rangle$ and $|2\rangle$), one can use different preparations of the initial populations to control the MI [see Fig.~\ref{fig3}(c)] and hence to create new optical patterns in the system.

\begin{figure*}
\centering
\includegraphics[width=0.96\textwidth]{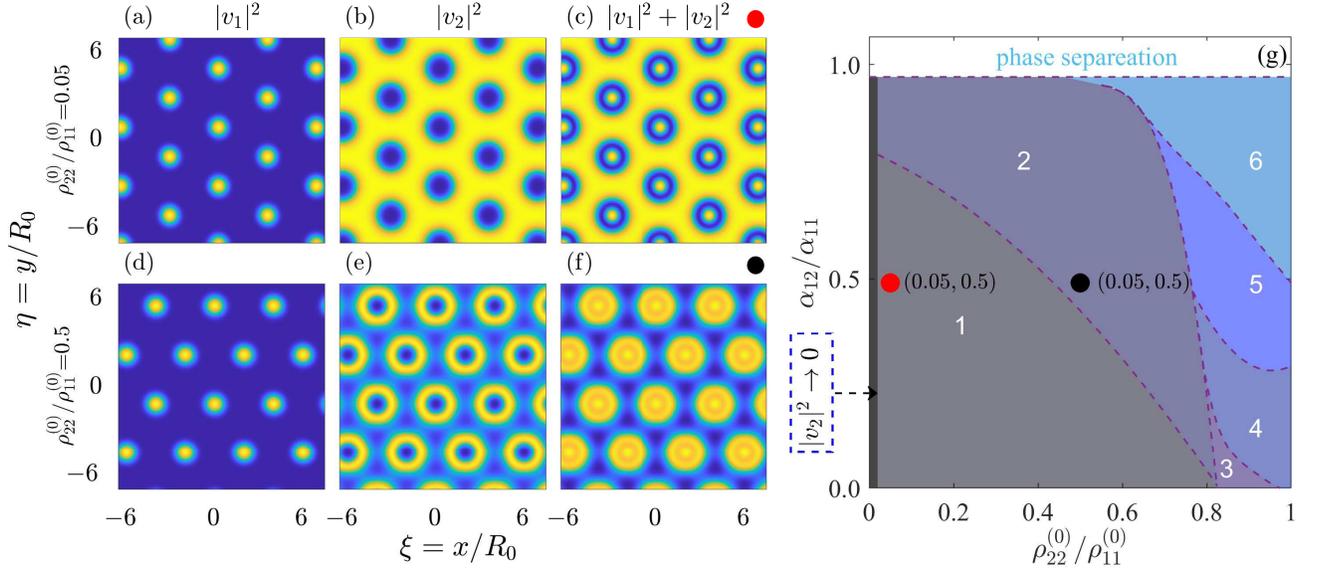}
\caption{\footnotesize Pattern formation of the two-component probe field controlled  by the ratio of initial populations in the two atomic ground states $|1\rangle$ and $|2\rangle$, i.e.  $\rho_{22}^{(0)}/\rho_{11}^{(0)}$, and by the ratio of nonlinearity  $\alpha_{12}/\alpha_{11}$, by taking the dimensionless normalized probe-field intensity $|v|^2=|v_1|^2+|v_2|^2$ as a function of the dimensionless coordinates $\xi = x/R_0$ and $\eta = y/R_0$~(with $\sigma=1.5$ and $\alpha_{11}=500$) for different $\rho_{22}^{(0)}/\rho_{11}^{(0)}$.
(a)-(c) Optical patterns of $|v_1|^2$, $|v_2|^2$, $|v_1|^2+|v_2|^2$ for $\rho_{22}^{(0)}/\rho_{11}^{(0)}=0.05$ and $\alpha_{12}/\alpha_{11}=0.5$;
(d)-(f) Similar to (a)-(c) but for $\rho_{22}^{(0)}/\rho_{11}^{(0)}=0.5$ and $\alpha_{12}/\alpha_{11}=0.5$.
(g)~Phase diagram for the optical pattern formation, in which different domains~(phases) are obtained by using different values
of ($\rho_{22}^{(0)}/\rho_{11}^{(0)},\alpha_{12}/\alpha_{11}$). The domain 1 and domain 2 correspond to the optical patterns illustrated in (c) and (f), respectively. The optical patterns corresponding to domains 3-6 are similar to those of Fig.~\ref{fig4}(b)-(e) (omitted here). The domain on the top of (g) (white color) is the one with phase separation; the domain on the leftmost side (black color) is the one where the $v_2$-component disappears.
}
\label{fig6}
\end{figure*}

We consider a general case by assuming the initial populations in the ground states  $|1\rangle$ and $|2\rangle$ can be adjusted arbitrarily (with $\sigma=1.5$ and $\alpha_{11}=500$ fixed). As done above, we look for the stationary states of the system through numerically solving Eq.~(\ref{cnls3}) under the condition of minimum energy.
Fig.~\ref{fig6} shows the optical patterns [Fig.~\ref{fig6}(a)-(f)] and the related phase diagram [Fig.~\ref{fig6}(g)] of the two-component probe field.

Illustrated in Figs.~\ref{fig6}(a)-(c) are respectively for the intensity distributions $|v_1|^2$, $|v_2|^2$, and $|v_1|^2+|v_2|^2$ as functions of $\xi=x/R_0$ and $\eta = y/R_0$  by taking $(\rho_{22}^{(0)}/\rho_{11}^{(0)},\alpha_{12}/\alpha_{11})=(0.05,0.5)$
(with the other parameters the same as those used in Fig.~\ref{fig5}). One sees that a new type of optical structure appears in the system. The pattern for the $v_1$-component is a hexagonal lattice consisting of bright circular spots [small, solid yellow circles; see Fig.~\ref{fig6}(a)]. While the pattern for the $v_2$-component is very different, in which there is a bright background with inserted low-intensity hexagonal optical spots [large, solid blue circles; see Fig.~\ref{fig6}(b)]. The superposition of the two components, i.e. $|v_1|^2+|v_2|^2$, displays as a hexagonal lattice with the spots of dark circular rings [Fig.~\ref{fig6}(c)]. This situation is relevant to the domain 1 in the phase diagram shown by Fig.~\ref{fig6}(g).

Another type of optical structure can be found when the population in the ground state $|2\rangle$ is increased, by taking $(\rho_{22}^{(0)}/\rho_{11}^{(0)},\alpha_{12}/\alpha_{11})=(0.5,0.5)$.
The optical patterns in this case for $|v_1|^2$, $|v_2|^2$, and $|v_1|^2+|v_2|^2$ are shown in Figs.~\ref{fig6}(d)-(f), respectively.
We see that the distribution of the $v_1$-component is a hexagonal lattice
consisting of small, solid yellow circles [Fig.~\ref{fig6}(d)]; the distribution of the $v_2$-component is also a hexagonal lattice, but the spots of the lattice are large, yellow circular rings [Fig.~\ref{fig6}(e)]. As a result, the superposition of the two components $|v_1|^2+|v_2|^2$ displays an interesting hexagonal lattice consisting of two-ring spots [Fig.~\ref{fig6}(f)].
The case considered here corresponds  to the domain 2 in the phase diagram shown by Fig.~\ref{fig6}(g).

We have made further simulations on the optical pattern formation in the system by choosing other initial populations in the atomic ground states $|1\rangle$ and $|2\rangle$, with the results given in the following:

(i)~When $\rho_{22}^{(0)}/\rho_{11}^{(0)}\sim 0$ (i.e. the initial  populations are nearly prepared in the ground state $|1\rangle$), we find that the optical pattern occurs only for the $v_1$-component, which has a hexagonal-shaped structure. This case corresponds to the thin domain with gray color shown in the leftmost side of Fig.~\ref{fig6}(g). This is easy to understand because in this situation the system is reduced into a ladder-shaped excitation scheme (i.e. the single Rydberg-EIT involved only the atomic states $|1\rangle$, $|3\rangle$, and $|4\rangle$) when $\rho_{22}^{(0)}/\rho_{11}^{(0)}\sim 0$. Similarly, if $\rho_{11}^{(0)}/\rho_{22}^{(0)}\sim 0$ the optical pattern occurs only for the $v_2$-component, which has also a hexagonal-shaped structure allowed by  the single Rydberg-EIT involved the atomic states $|2\rangle$, $|3\rangle$, and $|4\rangle$.

(ii)~The two types of optical lattice patterns shown in Fig.~\ref{fig6}(c) and Fig.~\ref{fig6}(f) can be converted each other through adjustments of $\rho_{22}^{(0)}/\rho_{11}^{(0)}$ and $\alpha_{12}/\alpha_{11}$.

(iii)~When the $\rho_{22}^{(0)}/\rho_{11}^{(0)}$ is close to 1, i.e. the initial populations in two ground states are approximately equal (e.g. $\rho_{22}^{(0)}/\rho_{11}^{(0)}>0.8$), the optical patterns are similar to those found in Fig.~\ref{fig5}(b)-(e), which are relevant to the domains 3-6 of Fig.~\ref{fig6}(g)~(not shown here for saving space).

In addition, an optical phase separation can occur for large $\alpha_{12}/\alpha_{11}$, as indicated by the white color domain on the top of Fig.~\ref{fig6}(g), which will be discussed in detail in the following section.

\subsection{Optical phase separations and their control}\label{sec3E}

\begin{figure*}[htpb]
\centering
\includegraphics[width=0.96\textwidth]{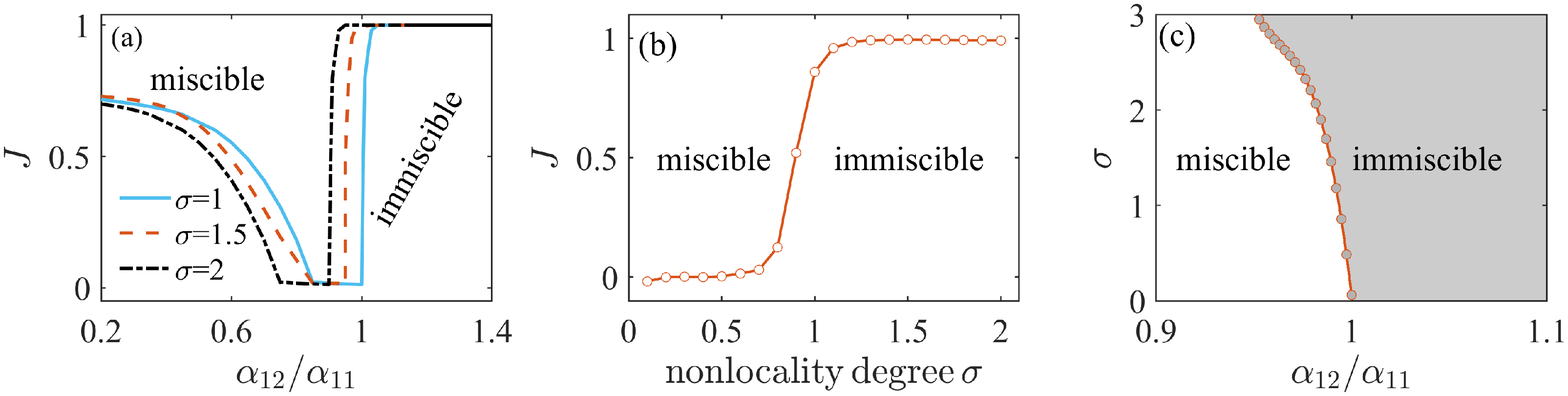}
\caption{\footnotesize  Separation degree describing the optical phase separation and the phase diagram describing the transition from miscible to immiscible states.
(a)~Separation degree $J$ as a function of the ratio of nonlinearity $\alpha_{12}/\alpha_{11}$ (with $\alpha_{11}=500$) and nonlocality degree of the Kerr nonlinearities $\sigma$. The solid blue, dashed red,
dotted black lines are for $\sigma=1$, 1.5, 2, respectively.
(b)~The circled solid red line is the separation degree $J$ as a function of $\sigma$ (with $\alpha_{11}=500$ and $\alpha_{12}/\alpha_{11}=0.95$).
(c)~Phase diagram describing the transition from the miscible to immiscible states in the parameter plane ($\alpha_{12}/\alpha_{11}$, $\sigma$).  The circled solid red line is the boundary of the transition;
the gray domain is relevant to Fig.~\ref{fig4}(f) and \ref{fig5}(f) where
optical phase separations occur.
}
\label{fig7}
\end{figure*}

We now investigate the optical phase separation in the system mentioned in the last three subsections in detail. Phase separation is a ubiquitous phenomenon in nature~\cite{Glotzer1994,Timmermans1998,Ao1998,Pooley2005,Sabbatini2011,Wen2012,
Shevtsova2015,Cates2018,Ota2019,Kumar2019}, which refers to a state of the system where two or more phases occur in different parts of the system, and the most common type of phase separation is between two immiscible fluids such as oil and water. We have shown that similar phenomenon may appear in the present Rydberg atomic system, as illustrated in Fig.~\ref{fig4}(f) and Fig.~\ref{fig5}(f) where the distributions of two polarization components of the probe field appear separately in space.

The physical reason for the appearance of such optical phase separation can be understood as follows. In many aspects, the motion of a light field behaves like that of a fluid flow~\cite{Carusotto2013}. In our system, the probe field has two components, which is similar to a mixture of two fluids. Since the interaction between Rydberg atoms under our study is repulsive, the system behaves like a two-fluid mixture and the two components of the probe field will be immiscible if the cross-Kerr nonlinearities (denoted by the nonlinearity parameter $\alpha_{12}$ which controls the mutual-interaction between the two components) reaches at a critical value.

From the phase diagrams shown in the panel (g) of Fig.~\ref{fig4}--Fig.~\ref{fig6}, we see that optical phase separations
occur principally for the case of $\alpha_{12}>\alpha_{11}$. This is easy to understand because for realizing the phase separations the mutual-interaction between the two probe components (characterized by the cross-Kerr nonlinearities) must be larger than the self-interaction (characterized by the self-Kerr nonlinearities) in the two probe components.

To quantify the optical phase separations in the system, we define the {\it separation degree}
\begin{align}\label{EqJ}
J=1-\frac{\int  |v_1(\vec{\zeta})||v_2(\vec{\zeta})|d^2\zeta}{\big[\int  |v_1(\vec{\zeta})|^2d^2\zeta\times \int  |v_2(\vec{\zeta})|^2d^2\zeta\big]^{1/2}}.
\end{align}
$J=0$~($J=1$) describes a complete spatial overlap--miscibility (complete spatial separation--immiscibility) of the two components of the probe field. If $0<J<1$, the two components have a mixture in space, with the degree of the separation determined by the value of $J$. For example, in panels (d), (e), and (f) of Fig.~5, the values of $J$ are $0.37$, $0$, and $1$, respectively. When $J$ has an intermediate value between 0 and 1 [e.g. Fig.~(5d) for which $J=0.37$], the intensity distributions of both the two polarization components are full of whole space, with the bright spots locating in different positions. This is different from the case for $J=1$ [i.e. Fig.~5(f)], where the distributions of the two polarization components are completely separated in space.

Shown in Fig.~\ref{fig7}(a) is the separation degree $J$ as a function of the ratio of nonlinearity $\alpha_{12}/\alpha_{11}$ (with  $\alpha_{11}=500$ fixed) for different nonlocality degree of the Kerr nonlinearities $\sigma$. The solid blue, dashed red, and dotted black lines in the figure are for $\sigma=1$, 1.5, and 2, respectively.
We see  that the two probe components are miscible (i.e. $J<1$) for smaller
$\alpha_{12}/\alpha_{11}$ but immiscible (i.e. $J=1$) for larger $\alpha_{12}/\alpha_{11}$, agreeing with the analysis given above and the results obtained in Fig.~\ref{fig4}(a)-(c) and Fig.~\ref{fig5}(a)-(c).

The circled solid red line in Fig.~\ref{fig7}(b) is the separation degree $J$ as a function of nonlocality degree $\sigma$ (with $\alpha_{11}=500$ and $\alpha_{12}/\alpha_{11}=0.95$ fixed). One sees that the two probe components are miscible when the nonlocality degree $\sigma$ is small, but they are immiscible when $\sigma$ is large. This shows that the nonlocality degree of the Kerr nonlinearities can be used to manipulate the optical phase separation in the system.

In order to find a criterion for the transition between the miscible and immiscible states, following Refs.~\cite{Ao1998,Kumar2019,Wen2012}, the expression of the energy difference between the miscible and immiscible states of the system can be derived, given by
\begin{align}\label{MIT0}
\Delta E=& \frac{I_1I_2}{V_{\rm tot}} \Big[
\int \frac{1}{2}\big[\mathcal{\Re}_{12}(\vec{\zeta})+\mathcal{\Re}_{21}(\vec{\zeta})\big] d^2\zeta\notag\\
&-\Big(\int \mathcal{\Re}_{11}(\vec{\zeta})d^2\zeta \times\int \mathcal{\Re}_{22}(\vec{\zeta})d^2\zeta\Big)^{1/2}\Big],
\end{align}
with the derivation presented in Appendix~\ref{appC}. Here $V_{\rm tot}=V_1+V_2$ with $V_1$ and $V_2$ the volumes occupied by $v_1$- and $v_2$-components, respectively.
If $\Delta E>0$ ($\Delta E<0$), i.e. the arithmetic mean of the nonlocal response functions for the cross-Kerr nonlinearities $\Re_{12}$ and $\Re_{21}$ is lager (smaller)
than the geometric mean of the response functions of the self-Kerr nonlinearities $\Re_{11}$ and $\Re_{22}$, the system will have higher~(lower) energy and hence tends to
go into immiscible (miscible) state.
The case $\Delta E =0$ defines the critical value~(boundary) for the transition from miscible to immiscible states~\cite{note3}.

Plotted in Fig.~\ref{fig7}(c) is the result on the phase diagram describing the transition from the miscible to immiscible states in the parameter plane ($\alpha_{12}/\alpha_{11}$, $\sigma$) by using the criterion (\ref{MIT0}).
The circled solid red line in the figure is the boundary for the transition;
the gray domain is corresponding to Fig.~\ref{fig4}(f) and \ref{fig5}(f) where optical phase separations occur.
We see that the domain for the optical immiscibility can be enlarged
by increasing the nonlocality degree of the Kerr nonlinearities $\sigma$,
which can be realized in the Rydberg atomic gas since the long-ranged Rydberg-Rydberg interaction plays an important role in such system.

\subsection{Nonlocal two-component optical solitons and vortices}\label{sec3F}

In all of the considerations given above, the system is assumed to work in regimes where the four nonlocal response functions $\Re_{jl}$ ($j,l=1,2$) in Eqs.~(\ref{cnls3}) are negative, based on which various optical patterns are obtained. Since the Rydberg atomic gas under consideration can be actively manipulated, one can choose different parameters to make the signs of $\Re_{jl}$ change, and hence it is possible to generate nonlocal two-component spatial solitons and vortices in the system.  For example, all the four response functions can be made to be positive if the system parameters are chosen to be
$\Delta_2=-2\pi\times 0.4\,{\rm MHz}$,
$\Delta_3=-2\pi\times3.0\,{\rm MHz}$,
${\rm \Delta_4=-2\pi\times 250\,MHz}$,
 ${\rm \Gamma_{12}=\Gamma_{21}=2\pi\times 6}$ MHz,
$\Gamma_3=\Gamma_{4}=2\pi\times 16.7$ kHz, $\Omega_{c}=2\pi\times 25\,{\rm MHz}$, and $\mathcal{N}_a=2.0\times 10^{11}\,{\rm
cm^{-3}}$.
In this situation, the interaction between the two probe components becomes attractive; a plane-wave state may undergo a MI, which does not result in the formation of (extended) optical patterns but (localized) bright-bright soliton pairs.
\begin{figure}[h]
\centering
\includegraphics[width=0.485\textwidth]{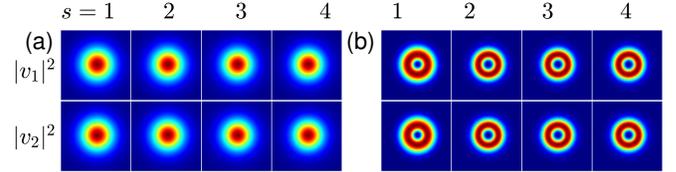}
\caption{\footnotesize
Formation of nonlocal spatial optical soliton and vortex pairs.
(a)~Spatial distributions of the bright-bright soliton pair for $s = z/(2L_{\rm diff})= 1,\,2,\,3,\,4$, by taking $|v_1|^2$ and $|v_2|^2$ as functions of $\xi = x/R_0$ and $\eta = y/R_0$ for $\sigma=1$.
(b)~Spatial distributions of the vortex-vortex pair for $s = 1,\,2,\,3,\,4$ for $\sigma=1.8$.
}
\label{fig8}
\end{figure}

Fig.~\ref{fig8}(a) shows the spatial distributions of a bright-bright soliton pair when it propagates respectively to the positions $s = z/(2L_{\rm diff})= 1,\,2,\,3,\,4$ by taking $|v_1|^2$ (top row) and $|v_2|^2$ (bottom row) as  functions of $\xi = x/R_0$ and $\eta = y/R_0$  for $\sigma=1$ (which gives $L_{\rm diff}=0.94$\,mm).
The result is obtained by numerically solving Eqs.~(\ref{cnls3}) based on the above parameters and with the initial condition $v_1=v_2=2\,{\rm sech}(\sqrt{\xi^2+\eta^2}/2)$. We see that the soliton pair is quite stable during propagation.

Illustrated in Fig.~\ref{fig8}(b) are the spatial distributions of a vortex-vortex pair when propagating respectively to the positions $s = z/(2L_{\rm diff})= 1,\,2,\,3,\,4$ for $\sigma=1.8$ (which gives $L_{\rm diff}=0.29$\,mm). The result is obtained also by numerically solving Eqs.~(\ref{cnls3}) with the initial condition $v_1=v_2=2.3\,\zeta\, L_0^{1}(-\frac{\xi^2+\eta^2}{2})\exp (-\frac{\xi^2+\eta^2}{4})e^{i\phi}$.
Here $L_p^{m}$ is generalised Laguerre polynomial with $(p,m)$ the mode indexes~\cite{Andrews2013}; $\phi=\arctan(\eta/\xi)$ is a phase. One sees that the vortex-vortex pair is also quite stable during propagation.

If other system parameters are suitably chosen,  one can realize the condition $\Re_{11}<0$, $\Re_{12}<0$, $\Re_{21}>0$, and $\Re_{22}>0$. In this case, the system supports stable bright-dark soliton pairs. The reason of the stability for both the 3D soliton-soliton and vortex-vortex pairs described above is due to fact that the Kerr nonlinearities in the Rydberg gas are nonlocal.

\section{Discussion and summary}\label{sec4}

The predictions of the stationary optical patterns and related structural phase transitions presented above may be observed experimentally by using a cold Rydberg atomic gas with the inverted Y-shaped level configuration. Since these lattice patterns form in the transverse $x$-$y$ plane and the typical lattice separation is around 7\,$\mu$m, one should prepare a Rydberg gas of transverse size of several tens of micrometers (e.g., $30\,\mu$m); furthermore, to form a clear optical pattern based on the MI starting from a plane-wave probe field plus a random initial condition, the longitudinal size of the gas should be larger than the nonlinearity length $L_{\rm non}$ and the diffraction length $L_{\rm diff}$, both of them are around 0.2\,mm [see Fig.~\ref{fig1}(c1)-(c3)], which can be realized by current experimental techniques~\cite{Weidemuller2016}. Comparably, the observation of the spatial optical solitons and vortices needs flexible conditions, for which the transverse size of  10 micrometers is enough because the typical transverse size of the solitons and vortices is only of several micrometers. To realize the predicted optical structures, one can inject a continuous-wave probe beam with two orthogonally polarized components; under the condition of the double EIT and by setting suitable system parameters, the probe beam will undergo MI and then be transformed into the spatial optical patterns, solitons, and vortices along the $z$-direction for several millimeters.

In the present work, as in Ref.~\cite{Henkel2010}, we have limited our considerations only for stationary optical structures based on the assumption that the probe field is a CW spatial light beam; the spatial optical structures are obtained by using an ``imaginary-time''  propagation method [i.e. $z \rightarrow iz$ in Eq.~(\ref{cnls00})], by which the ground-state solutions can be acquired very quickly for short propagation distance. A canonical approach for the optical pattern formation relevant to experimental observation
would be to propagate the light fields through the medium along $z$ direction and look at the transverse patterns after some propagation distance by using a real-time propagation method [which has been used only for Fig.~\ref{fig1}(c) in our work]. It is available to get temporally oscillatory optical structures in the system through Hopf bifurcations by considering the case beyond CW approximation,
so that the eigen values of the eigen equations in MI analysis may have both real and imaginary parts (such as those described in Refs.~\cite{Scroggie1994,Tlidi1997,Tlidi2003}), an interesting topic deserving to be explored further.

In conclusion, we have investigated the formation and control of spatial optical patterns in a cold Rydberg atomic gas via double Rydberg-EIT. Based on the analysis on the coupled 3D NNLS equations derived from the MB equations, we found that through the MI of the homogeneous plane-wave state of the two-component probe field, the system undergoes a spontaneous symmetry breaking and hence the emerge of various self-organized optical lattice structures, which can be controlled by the ratio between the cross- and self-Kerr nonlinearities, the nonlocality degree of the Kerr nonlinearities, and the populations initially prepared in the two atomic ground states. In particular, a crossover from phase mixture to phase separation of the two components of the probe field may appear when the ratio between the cross- and self-Kerr nonlinearities passes over a critical value. We also found that the system supports nonlocal two-component spatial optical soliton and vortices when the parameters of the system are selected suitably.
The results obtained in this work are not only useful for obtaining novel optical structures and realizing their active manipulations, but also promising for practical applications, e.g. for the design of nonlinear optical splitters based on the optical phase separation reported here.

\section*{Acknowledgments}

This work was supported by the National Natural Science Foundation of China under Grant No.~11975098 and the Research Funds of Happiness Flower ECNU under Grant No.~2020ECNU-XFZH005.

\appendix
\begin{widetext}
\section{Optical Bloch equation and its solutions}\label{appA}
\subsection{Optical Bloch equation}
The explicit expression of the optical Bloch equation (\ref{Bloch0}) is given by~\cite{Mu2021}
\begin{subequations}\label{DMs1}
\begin{align}
&i\frac{\partial \rho _{11}}{\partial t}+i\Gamma_{21}\rho_{11}-i\Gamma_{12} \rho _{22}-i\Gamma_{13} \rho_{33}+\Omega ^*_{p1}\rho _{31}
-\Omega_{p1}\rho_{13}=0, \\
&i\frac{\partial \rho_{22}}{\partial t}-i\Gamma_{21}\rho_{11}+i\Gamma_{12} \rho_{22}-i\Gamma_{23} \rho_{33}+\Omega^*_{p2}\rho_{32}-\Omega_{p2}\rho _{23}=0 , \\
&i\frac{\partial \rho_{33}}{\partial t}-i\Gamma_{34} \rho _{44}+i\Gamma_{3}\rho_{33}+(\Omega_{p1} \rho_{13}%
+\Omega_{p2}\rho_{23}+\Omega^*_c \rho_{43}-{\rm c.c.})=0, \\
&i \frac{\partial \rho_{44}}{\partial t}+i \Gamma_{34} \rho_{44}-\Omega^*_c \rho_{43}+\Omega_c \rho_{34}=0,
\end{align}
\end{subequations}
for diagonal elements, and
\begin{subequations}\label{DMs2}
\begin{align}
&\Big(i\frac{\partial }{\partial t}+d_{21}\Big)\rho_{21}+\Omega _{p2}^{*}\rho_{31}-\Omega _{p1}\rho_{23}=0,\\
&\Big(i\frac{\partial }{\partial t}+d_{31}\Big)\rho _{31}+\Omega _{c}^{*}\rho _{41}+\Omega _{p1}(\rho _{11}-\rho_{33})+\Omega_{p2}\rho_{21}=0, \\
&\Big(i\frac{\partial }{\partial t}+d_{32}\Big)\rho _{32}+\Omega _{c}^{*}\rho _{42}+\Omega _{p1}\rho _{12}+\Omega_{p2}(\rho_{22}-\rho_{33})=0, \\
&\Big(i\frac{\partial }{\partial t}+d_{41}\Big)\rho _{41}+\Omega _{c}\rho _{31}-\Omega _{p1}\rho _{43}
-\frac{1}{2}\mathcal{N}_a\int d^{3}{ r^{\prime}}  \mathcal{V}({\bf r^{\prime}}-{\bf r})  \rho_{44,41}({\bf r^{\prime}},{\bf r},t)=0,\\
&\Big(i\frac{\partial }{\partial t}+d_{42}\Big)\rho _{42}+\Omega _{c}\rho _{32}-\Omega _{p2}\rho _{43}
-\frac{1}{2}\mathcal{N}_a\int d^{3}{ r^{\prime}}  \mathcal{V}({\bf r^{\prime}}-{\bf r}) \rho_{44,42}({\bf r^{\prime}},{\bf r},t)=0,\\
&\Big(i\frac{\partial }{\partial t}+d_{43}\Big)\rho _{43}-\Omega ^*_{p1}\rho _{41}-\Omega^*_{p2}\rho_{42}+\Omega _{c}(\rho _{33}-\rho_{44})-\frac{1}{2}\mathcal{N}_a\int d^{3}{r^{\prime}}  \mathcal{V}({\bf r}^{\prime}-{\bf r}) \rho_{44,43}({\bf r^{\prime}},{\bf r},t)=0,
\end{align}
\end{subequations}
\end{widetext}
for non-diagonal elements. Here $d^3r'= dx'dy'dz'$;   $d_{\alpha\beta}=\Delta_{\alpha}-\Delta_{\beta}+i \gamma_{\alpha\beta}$ ($\alpha,\beta=1,2,3,4; \alpha \neq  \beta$); $\Delta_1=0$, $\Delta_2=\omega_1-\omega_2$, $\Delta_3=\omega_{p}-(\omega_3-\omega_1)$, and  $\Delta_4=\omega_c+\omega_{p}-(\omega_4-\omega_1)$
are respectively one- and  two-photon detunings;
$\gamma_{\alpha\beta}=(\Gamma_{\alpha}+\Gamma_{\beta})/2
+\gamma_{\alpha\beta}^{\rm dep}$ with $\Gamma_{\alpha}=\sum_{\alpha<\beta}\Gamma_{\alpha\beta}$  [$\Gamma_{\alpha\beta}$ and $\gamma_{\alpha\beta}^{\rm dep}$ are the spontaneous emission decay rate and dephasing rate from $|\beta\rangle$ to $|\alpha\rangle$, respectively].

Although the Eq.~(\ref{Bloch0}) describe the evolution of the matrix elements $\rho_{\alpha\beta}({\bf r},t)$ of the one-body reduced DM ${\rho} ({\bf r},t)$, they involve two-body reduced DM  ${\rho}_{\rm twobody} (\bf r^{\prime},{\bf r},t)$ with matrix elements defined by $\rho_{\alpha\beta,\mu\nu}({\bf r^{\prime},r},t)=\langle
\hat{S}_{\alpha\beta}({\bf r^{\prime}},t) \hat{S}_{\mu\nu}({\bf r},t)\rangle$  (or called two-body correlators) due to the Rydberg-Rydberg interaction.
So Eq.~(\ref{Bloch0}) is not a closed one and one must solve the equations for the two-body DM elements. For example, the equation of two-body matrix element $\rho_{44,41}\equiv \langle\hat{S}_{44}({\bf r^{\prime}},t) \hat{S}_{41}({\bf r},t)\rangle$ reads
{\small \begin{eqnarray}\label{2body}
&&\Big[i \frac{\partial}{\partial t}+d_{41}+i \Gamma_{34}-\mathcal{V}\left(\mathbf{r}^{\prime}-\mathbf{r}\right)\Big]
\rho_{44,41}\notag\\
&&-\Omega_{c}^{*}\rho_{43,41}+\Omega_{c}(\rho_{34,41}+\rho_{44,31})-\Omega_{p 1}\rho_{44,43}\notag\\
&& -\frac{\mathcal{N}_{a}}{2}\int d^3 r^{\prime \prime} \mathcal{V}\left(\mathbf{r}^{\prime \prime}-\mathbf{r}\right)\rho_{44,44,41}(\mathbf{r}'',\mathbf{r}',\mathbf{r}, t)=0.
\end{eqnarray}}\noindent
We see that it involves three-body DM element $\rho_{44,44,41}(\mathbf{r}'',\mathbf{r}',\mathbf{r},t) \equiv\langle\hat{S}_{44}(\mathbf{r}'',t) \hat{S}_{44}(\mathbf{r}',t)
\hat{S}_{41}(\mathbf{r},t)\rangle$. The equations for other two-body DM elements are too lengthy and hence omitted here for saving space. Thus, to solve the equations of the one-body and two-body DM elements above we need the equations of three-body DM elements. In a similar way, we can get the equations of motion for the three-body DM elements, which, however, involve four-body DM elements (omitted here), and so on. To solve such a chain of infinite equations for $N$-body ($N=1,2,3,...$) DM elements, a sutiable truncation technique beyond mean-field approximation is needed, which has been developed recently~\cite{Bai2016,Zhang2018,Bai2019}. Based on such approach, we can solve the MB equations (\ref{Bloch0}) and (\ref{Maxwell}) by using an asymptotic expansion in a consistent and standard way.

\subsection{Equations of motion of density-matrix elements and their solutions up to third-order approximations}

We are interested in stationary states of the system, and hence the time derivatives in the MB equations (\ref{Bloch0}) can be neglected (i.e.,~$\partial/\partial t= 0$), which is valid if the probe and control fields have a large time duration.
We adopt the method developed in Refs.~\cite{Bai2016,Zhang2018,Bai2019} to firstly solve the Bloch equation (\ref{Bloch0}) under the condition of Rydberg-EIT. We assume that all the atoms are initially prepared in the ground states $|1\rangle$ and $|2\rangle$.

Since the probe field is weak, we can take  $\Omega_{pj}\sim \epsilon$ as expansion parameter, and make the expansion
$\rho_{\alpha\beta}=\epsilon \rho _{\alpha\beta}^{\left( 1
\right)}+\epsilon ^2\rho _{\alpha\beta}^{\left( 2 \right)}+\cdots$, $\left(\beta=1,2;\alpha=2,3,4;\beta<\alpha \right)$.
Substituting the expansions into the Eq.~(\ref{Bloch0}), and collecting coefficients of $\epsilon^l~ (l=0,1,2,\cdots)$, we obtain a series of linear but inhomogeneous equations that can be solved  order by order.

\vspace{2mm}
{\it (i) First-order solution:}
At the first order~$(l=1)$, we obtain
\begin{align}
&\rho _{i j}^{\left( 1 \right)}=[\delta_{i 3}d_{4j}-\delta_{i 4}\Omega _c]\rho_{jj}^{(0)}\Omega_{pj}/D_j,
\end{align}
for $i=3,4$ and $j=1,2$ (other $\rho_{\alpha\beta}^{(1)}$ are zero), with $D_j=\left|\Omega _c \right|^2-d_{3j} d_{4j}$.

\vspace{2mm}
{\it (ii) Second-order solution:}
Solving the second order~$(l=2)$ equations yields
\begin{subequations}
\begin{align}	
&\rho _{21}^{\left( 2 \right)}=a_{21}^{\left( 2 \right)}\Omega_{p1} \Omega_{p2}^* ,\\
&\rho _{\alpha\alpha}^{\left( 2 \right)}=a_{\alpha\alpha,1}^{\left( 2 \right)}|\Omega_{p1} |^2+a_{\alpha\alpha,2}^{\left( 2 \right)}|\Omega_{p2} |^2,\,(\alpha=1-4)\\
&\rho _{43}^{\left( 2 \right)}=a_{43,1}^{\left( 2 \right)}|\Omega_{p1} |^2+a_{43,2}^{\left( 2 \right)}|\Omega_{p2}|^2,
\end{align}
\end{subequations}
in which
\begin{align*}
&a_{33,j}^{(2)} =\rho_{jj}^{(0)} {N_j}/(i\Gamma_3),\\
&a _{21}^{\left( 2 \right)}=(\rho_{22}^{(0)} d_{42}^{*}/D_{2}^{*}-\rho_{11}^{(0)} d_{41}/{D_1})/ d_{21},\\
&a_{44,j}^{\left( 2 \right)} = \rho_{jj}^{(0)} |{\Omega _c}|^2\big[{M_j} + {{N_j}{M_3}}/(i\Gamma_3)\big]/({i\Gamma _{34} + M_3| \Omega _c |^2}),\\
&a_{22,j}^{(2)} = -\big[{{({\Gamma _{13}}+{\Gamma _{21}})a_{33,j}^{(2)} + {\Gamma _{21}}a_{44,j}^{(2)} + \delta_{1j}i{N_1}\rho_{11}^{(0)}}}\big]/\Gamma_2,\\
&a_{43,j}^{\left( 2 \right)} =  - \Omega _c\big[\rho_{jj}^{(0)} /D_j + a_{33,j}^{(2)} - a_{44,j}^{(2)}\big]/d_{43},\\
&a_{11,j}^{\left( 2 \right)} =  -\left (a_{22,j}^{(2)} + a_{33,j}^{(2)} + a_{44,j}^{(2)}\right).
\end{align*}
Here $N_{j}=2i\text {Im}(d_{4j}/D_j)$, $M_{j}=2i\text{Im}[1/(D_{j}^* d_{43}^*)]$, $M_3=2i\text{Im}(1/d_{43}^*)$, $\Gamma_2=\Gamma _{12} +\Gamma _{21}$, and $\Gamma_3= {\Gamma _{13}} + {\Gamma _{23}}~(j=1,2)$.

{\it (iii) Third-order solution:}
The solutions of $\rho_{\alpha 1}^{(3)}$ and $\rho_{\alpha 2}^{(3)}$~($\alpha=3,4$) are given by
{\small
\begin{subequations}\label{S3rd}
\begin{align}
\rho _{\alpha 1}^{(3)} &= \Big[(a_{\alpha 1,1}^{(3)}|\Omega_{p1}|^2 + a_{\alpha 1,2}^{(3)}|\Omega_{p2}|^2\Big]\Omega_{p1}
+{\mathcal{N}_a}\Omega_c^*/{D_1}\notag\\
&\times\int {{d^3}{ r}'{\mathcal{ V}} ({\bf  r}' - {\bf  r}} )\Big[a_{44,41,1}^{(3)}{\left| \Omega_{p1}\right|^2}+a_{44,41,2}^{(3)}{\left|\Omega_{p2} \right|^2}\Big]\Omega_{p1},\\
\rho _{\alpha 2}^{(3)} &=\Big[a_{\alpha 2,1}^{(3)}{\left|\Omega_{p1} \right|^2}+ a_{\alpha 2,2}^{(3)}{\left| \Omega_{p2} \right|^2}\Big]\Omega_{p2}
+ {\mathcal{N}_a}{\Omega_c^*}/{D_2}\notag\\
&\times \int {{d^3}{ r}'\mathcal{ V} ({\bf  r}' - {\bf  r}} )\Big[a_{44,42,1}^{(3)}{\left| \Omega_{p1} \right|^2}+a_{44,42,2}^{(3)}{\left| \Omega_{p2} \right|^2}\Big]\Omega_{p2},
\end{align}
\end{subequations}}\noindent
in which
\begin{align*}
&a_{31,j}^{(3)}=\big[{\Omega_c^* a_{43,j}^{(2)}+d_{41}(a_{11,j}^{(2)}-a_{33,j}^{(2)})+\delta_{2j}a_{21}^{(2)}d_{41}}
\big]/{D_1},\\
&a_{32,j}^{(3)}=\big[\Omega_c^*a_{43,j}^{(2)}+{d_{42}( a_{22,j}^{(2)}-a_{33,j}^{(2)})+\delta_{1j}a_{21}^{*(2)}d_{42}}\big]/{D_2},\\
&a_{41,j}^{(3)}=\big[{\Omega_c (a_{33,j}^{(2)}-a_{11,j}^{(2)})-d_{31} a_{43,j}^{(2)}-\delta_{2j}\Omega_c a_{21}^{(2)}}\big]/{D_1},\\
&a_{42,j}^{(3)}=\big[{\Omega_c( a_{33,j}^{(2)}-a_{22,j}^{(2)})-d_{32}a_{43,j}^{(2)}-\delta_{1j}\Omega_ca_{21}^{*(2)}}\big]/{D_2},
\end{align*}
where the coefficients of two-body DM elements $a_{44,4j,\alpha}^{(3)}=a_{44,4j,\alpha}^{(3)}({\bf r^{\prime}},{\bf r},t)$~($j=1,2;\alpha=1,2$) is the function of ${\bf r^{\prime},r}$ and $t$, yet to be determined (see below).

\vspace{2mm}
{\it (iv) Second-order solution for two-body DM elements}\\
Notice that for obtaining the solution of $\rho_{\alpha\beta}^{(3)}$~($\alpha=3,4;\beta=1,2$),
equations for some two-body DM elements $\rho_{\alpha\beta,\mu\nu}$ must be solved simultaneously. These two-body DM elements are nonzero starting at $\epsilon^2$-order, so they can be assumed to have the form $\rho_{\alpha\beta,\mu\nu}=\epsilon^2 \rho_{\alpha\beta,\mu\nu}^{(2)}+\epsilon^3 \rho_{\alpha\beta,\mu\nu}^{(3)}+\cdots$. Then we have the equations for the second-order two-body DM elements $\rho_{\alpha j,\beta j}^{(2)}$~($\alpha,\beta=3,4;\, j= 1,2$):
{\small
\begin{align}
&\left(\begin{array}{ccc}
d_{3j} & \Omega_{c}^{*} & 0 \\
\Omega_{c} & d_{3j}+d_{4j} & \Omega_{c}^{*} \\
0 & 2 \Omega_{c} & A
\end{array}\right)\left(\begin{array}{c}
\rho_{3j,3j}^{(2)} \\
\rho_{4j,3j}^{(2)} \\
\rho_{4j,4j}^{(2)}
\end{array}\right)=-\left(\begin{array}{c}
a_{3j}^{(1)} \\
a_{4j}^{(1)}  \\
0
\end{array}\right)\rho_{jj}^{(0)}\Omega_{pj}^{2},\\
&\left(\begin{array}{*{4}{c}}
	B&\Omega _c&\Omega _c&0 \\
	\Omega _c^*&d_{41} + d_{32}&0&\Omega _c \\
	\Omega _c^*&0&d_{42} + d_{31}&\Omega _c \\
	0&{\Omega _c^*}&{\Omega _c^*}&d_{32} + d_{31}
	\end{array} \right)
\left( {\begin{array}{*{20}{c}}
	{\rho _{42,41}^{\left( 2 \right)}} \\
	{ \rho _{41,32}^{\left( 2 \right)}} \\
	{ \rho _{42,31}^{\left( 2 \right)}} \\
	{ \rho _{32,31}^{\left( 2 \right)}}
	\end{array}} \right)\notag\\
&   = -\left( {\begin{array}{*{20}{c}}
	0 \\
	  a_{41}^{(1)}\rho_{22}^{(0)}\\
	{a_{42}^{(1)}}\rho_{11}^{(0)} \\
	{ a_{32}^{(1)}\rho_{11}^{(0)}}+a_{31}^{(1)}\rho_{22}^{(0)}
	\end{array}} \right)\Omega_{p1}\Omega_{p2},
\end{align}}\noindent
with $A=2 d_{4j}-\frac{1}{2}\mathcal{V}(\bf r)$  and $B=d_{41} + d_{42}-\frac{1}{2}\mathcal{V}(\bf r)$. We obtain $\rho_{\alpha j,\beta j}^{(2)}={Q_{0j}}/[{P_{0j}+P_{1j}{\cal V}(\bf r)}]$, where $P_{0j},P_{1j},Q_{0j}$ are functions of the detunings, spontaneous emission decay rate, and dephasing rates of the system (their explicit expressions are omitted here).

\vspace{2mm}
{\it (v) Third-order solution for two-body DM elements}\\
The third-order solution of the two-body DM elements satisfies the equation
\begin{align}\label{rho3a}
{\bf M}\cdot \mathcal{\rho}= \mathbf{C}_1|\Omega_{p2}|^2\Omega_{p1}+\mathbf{C}_2|\Omega_{p1}|^2\Omega_{p1},
\end{align}
where{\small
\begin{align}
&{\bf M}=\left(
\begin{matrix}
m_{1}              & -\Omega _{c}^*&\Omega _{c}&\Omega _{c}&0          &0            &0        &0\\
-\Omega _c     &m_{2}              &0          &0          &\Omega _{c}&\Omega _{c}  &0        &0\\
-\Omega _{c}^* &0			   &m_{3}          &0          &\Omega_c^* &0            &-\Omega_c&0\\
\Omega _{c}^*  &0              &0		   &m_{4}		   &0          &-\Omega _c^* &\Omega_c &0\\
-i\Gamma_{34}  &\Omega_c^*	   &-\Omega_c  &0          &m_{5}          &0            &0        &\Omega_c\\
0			   &\Omega_c^*     & 0         &-\Omega_c  &0          &m_{6}           &0		   &\Omega_c\\
0              &0              &-\Omega_c^*&-\Omega_c^*&0          &0            &m_{7}       &\Omega_c^*\\
0              &0              &0          &-i\Gamma_{34}&\Omega_c^*&\Omega_c^*  &-\Omega_c&m_{8}\\
\end{matrix}
\right),\notag\\
& \mathbf{C}_1=[0,
 a_{42,41}^{(2)},
 a_{42,14}^{*(2)},
 -a_{44,2}^{(2)}\rho_{11}^{(0)},
 a_{41,32}^{(2)}-a_{41,23}^{(2)},
 a_{42,31}^{(2)}\notag\\&-a_{43,2}^{(2)}\rho_{11}^{(0)},
 a_{42,13}^{*(2)}+a_{43,2}^{*(2)}\rho_{11}^{(0)},
 a_{32,31}^{(2)}-a_{33,2}^{(2)}\rho_{11}^{(0)}-a_{32,13}^{*(2)}]^T,\notag\\
 &\mathbf{C}_2=[0,
 a_{41,41}^{(2)},
 a_{41,14}^{*(2)},
 -a_{44,1}^{(2)}\rho_{11}^{(0)},
 a_{41,31}^{(2)}-a_{41,13}^{(2)},
 a_{41,31}^{(2)}\notag\\&-a_{43,1}^{(2)}\rho_{11}^{(0)},
 a_{41,13}^{*(2)}+a_{43,1}^{*(2)}\rho_{11}^{(0)},
 a_{31,31}^{(2)}-a_{31,13}^{(2)}-a_{33,1}^{(2)}\rho_{11}^{(0)}
 ]^T,\notag\\
 &\rho=\big[
 \rho _{44,41}^{\left( 3 \right)},
 \rho _{43,41}^{\left( 3 \right)},
 \rho _{43,14}^{*\left( 3 \right)},
 \rho _{44,31}^{\left( 3 \right)},
 \rho _{41,33}^{\left( 3 \right)},
 \rho _{43,31}^{\left( 3 \right)},
 \rho _{43,13}^{*\left( 3 \right)},
 \rho _{33,31}^{\left( 3 \right)}\big]^{T},\notag
\end{align}}\noindent
with $m_{1}=i\Gamma_{34}+d_{41}-\frac{1}{2}\mathcal{V}$, $m_{2}=d_{43}+d_{41}-\frac{1}{2}\mathcal{V}$, $m_{3}=d_{43}^{*}+d_{14}^{*}$,
$m_{4}=i\Gamma_{34}+d_{31}$, $m_{5}=d_{41}+i\Gamma_{3}$, $m_{6}=d_{43}+d_{31}$,
$m_{7}=d_{43}^{*}+d_{13}^{*}$, and $m_{8}=i\Gamma_{3}+d_{31}$.

The equation for $\rho\rho_{44,42}^{(3)}$ reads
\begin{align}\label{rho3b}
{\bf N}\cdot \mathcal{\rho}= \mathbf{C}_3|\Omega_{p2}|^2\Omega_{p2}+\mathbf{C}_4|\Omega_{p1}|^2\Omega_{p2},
\end{align}
where
{\small
\begin{align}
&{\bf N}=\left(
\begin{matrix}
n_{1}             & -\Omega _{c}^*&\Omega _{c}&\Omega _{c}&0          &0            &0        &0\\
-\Omega _c     &n_{2}             &0          &0          &\Omega _{c}&\Omega _{c}  &0        &0\\
-\Omega _{c}^* &0			   &n_{3}          &0          &0	       &\Omega_c^*   &-\Omega_c&0\\
\Omega _{c}^*  &0              &0		   &n_{4}		   &-\Omega _c^*&0           &\Omega_c &0\\
0              &\Omega_c^*	   &0          &-\Omega_c  &n_{5}          &0            &0        &\Omega_c\\
-i\Gamma_{34}  &\Omega_c^*     &-\Omega_c &0          &0          &n_{6}            &0		   &\Omega_c\\
0			   &0              &-\Omega_c^* &-\Omega_c^*&0          &0            &n_{7}        &\Omega_c^*\\
0              &0              &0          &-i\Gamma_{34}&\Omega_c^*&\Omega_c^*  &-\Omega_c&n_{8}\\
\end{matrix}
\right)\notag\\
&\mathbf{C}_3=[0,
 a_{42,42}^{(2)},
 a_{42,24}^{(2)},
 -a_{44,2}^{(2)}\rho_{22}^{(0)},
 a_{42,32}^{(2)}-a_{43,2}^{(2)}\rho_{22}^{(0)},
 a_{42,32}^{(2)}\notag\\&-a_{42,23}^{(2)},
 a_{42,23}^{*(2)}+a_{43,2}^{*(2)}\rho_{22}^{(0)},
 a_{32,32}^{(2)}-a_{32,23}^{(2)}-a_{33,2}^{(2)}\rho_{22}^{(0)}
 ]^T,\notag\\
&\mathbf{C}_4=[0,
a_{42,41}^{(2)},
a_{42,14}^{(2)},
-a_{44,1}^{(2)},
a_{41,32}^{(2)}-a_{43,1}^{(2)},
a_{42,31}^{(2)}\notag\\&-a_{42,13}^{(2)},
a_{41,23}^{*(2)}+a_{43,1}^{*(2)}\rho_{22}^{(0)},
a_{32,31}^{(2)}-a_{32,13}^{(2)}-a_{33,1}^{(2)}\rho_{22}^{(0)}
]^T\notag\\
&\rho=[
\rho _{44,42}^{(3)},
\rho _{43,42}^{(3)},
 \rho _{43,24}^{*(3)},
 \rho _{44,32}^{(3)},
 \rho _{43,32}^{(3)},
 \rho _{42,33}^{(3)},
 \rho _{43,23}^{*(3)},
\rho _{33,32}^{(3)}],\notag
\end{align}}\noindent
with $n_{1}=i\Gamma_{34}+d_{42}-\frac{1}{2}\mathcal{V}$,
$n_{2}=d_{43}+d_{42}-\frac{1}{2}\mathcal{V}$,
$n_{3}=d_{43}^{*}+d_{24}^{*}$,
$n_{4}=i\Gamma_{34}+d_{32}$,
$n_{5}=d_{43}+d_{32}$,
$n_{6}=d_{42}+i\Gamma_{3}$,
$n_{7}=d_{43}^{*}+d_{23}^{*}$,
and $n_{8}=i\Gamma_{3}+d_{32}$.

By solving Eqs.~(\ref{rho3a}) and (\ref{rho3b}), we obtain the solution  $\rho_{44,4j}^{(3)}$ with the form
\begin{subequations}\label{twobody}
\begin{align}
&\rho_{44,41}^{(3)}=a_{44,41,1}^{(3)}\left|\Omega_{p 1}\right|^{2} \Omega_{p 1}+a_{44,41,2}^{(3)}\left|\Omega_{p 2}\right|^{2} \Omega_{p 1}, \\
&\rho_{44,42}^{(3)}=a_{44,42,1}^{(3)}\left|\Omega_{p 1}\right|^{2} \Omega_{p 2}+a_{44,42,2}^{(3)}\left|\Omega_{p 2}\right|^{2} \Omega_{p 2},
\end{align}
\end{subequations}
in which
\begin{equation}
a_{44,4 j, l}^{(3)}=\frac{\sum_{m=0}^2P_{j l m} \mathcal{V}^m\left(\bf{r}^{\prime}-\bf{r}\right)}{\sum_{n=0}^3Q_{j l n} \mathcal{V}^n\left(\bf{r}^{\prime}-\bf{r}\right)}.
\end{equation}
Here $P_{jln}$ and $Q_{jln}~(j, l = 1, 2) $ are constants, depending on the
spontaneous emission and dephasing rates, detunings, half Rabi frequency of the control field, as well as other parameters of the system.  Thereby, based on the first-, second-, and third-order solutions given above, one can obtain the explicit expression of the one-body DM elements up to third-order approximation, i.e. $\rho_{3j}\approx\rho_{3j}^{(1)}+\rho_{3j}^{(2)}+\rho_{3j}^{(3)}$~($j=1,2$).

\section{Expressions of the nonlinear optical susceptibilities and the derivation of the coupled NLSEs}\label{appB}

The optical susceptibility $\chi_{j}$  of the $j$th polarization component of the probe field is given by
$\chi_{j}=\mathcal{N}_a({\hat{\bf e}_{p\pm}\cdot{\bf p}_{j3})\rho_{3j}}/(\varepsilon_0{\cal E}_{\pm})$
($j=1,2$), which has the form
{\small
\begin{align}
\chi_{j}&=\chi_{j}^{(1)}
             +\Big[\chi_{j,{\rm loc} }^{(3,s)}+\chi_{j,{\rm nloc}}^{(3,s)}\Big]|{\cal E}_{p\pm}|^{2}
             +\Big[\chi_{j,{\rm loc}}^{(3,c)}+\chi_{j,{\rm nloc}}^{(3,c)}\Big]|{\cal E}_{p\mp}|^{2},
\end{align}}\noindent
where
{\small
\begin{subequations}\label{chip13}
\begin{align}
&\chi _{j}^{(1)}= \mathcal{N}_a| p_{j3} |^2 d_{4j}\rho_{jj}^{(0)}/(\varepsilon _0\hbar D_j), \label{chip1_1}\\
&\chi _{j,{\rm loc}}^{(3,s)}= \mathcal{N}_a a_{3j,j}^{(3)}| p_{j3} |^4/(\varepsilon _0\hbar ^3),\label{chip1_31}\\
&\chi _{j,{\rm loc}}^{(3,c)} = \mathcal{N}_aa_{3j,3-j}^{(3)}|p_{13}|^2|p_{23}|^2/(\varepsilon _0\hbar ^3),\label{chip1_32}\\
&\chi _{j,{\rm nloc}}^{(3,s)}=\frac{{\mathcal{N}_a^2}\Omega_c^*}{{{2\varepsilon _0}{\hbar ^3{D_j}}}}\int {{d^3}{ r}'\mathcal{V} ({\bf r}' - {\bf r})}a_{44,4j,j}^{(3)}({\bf r}',{\bf r}){| {{{\bf p}_{j3}}}|^4},\label{chip1_33}\\
&\chi _{j,{\rm nloc}}^{(3,c)}=\frac{\mathcal{N}_a^2\Omega_c^*}{2\varepsilon _0\hbar ^3D_j}\int d^3 r'\mathcal{V} ({\bf r}' - {\bf r}) a_{44,4j,3-j}^{(3)}({\bf r}',{\bf r})| {\bf p}_{13}|^2| {\bf p}_{23}|^2.\label{chip1_34}
\end{align}
\end{subequations}}\noindent

Substituting the results of $\rho_{31}$ and $\rho_{32}$~[Eqs.~(\ref{S3rd}) and (\ref{twobody})]
into Maxwell Eq.~(\ref{Maxwell}), we obtain the envelope equations controlling the dynamics of
the two polarization components of the probe field, which have the form of the coupled 3D NNLS equations
{\small
\begin{subequations}\label{cnls1}
\begin{align}
i\frac{{\partial {\Omega_{p1}}}}{{\partial z}} &+ \frac{c}{{2{\omega _{p}}}}\nabla _ \bot ^2\Omega_{p1}+ ({W_{11}}{\left| \Omega_{p1} \right|^2} + {W_{12}}{\left| \Omega_{p2}\right|^2})\Omega_{p1} \notag \\
&+ \int {d^3r' \mathcal{N}_{11}({\bf r'} -{\bf  r}){{\left| {{\Omega_{p1}({\bf r'})}} \right|}^2} \Omega_{p1}({\bf r})}\notag \\
&+ \int {{d^3}{ r'} {\mathcal{N}_{12}({\bf r'} -{\bf  r})}{{\left| {\Omega_{p2}({\bf r'})} \right|}^2}{\Omega_{p1}({\bf r})}} = 0, \\
i\frac{{\partial {\Omega_{p2}}}}{{\partial z}}  &+ \frac{c}{{2{\omega _{p}}}}\nabla _ \bot ^2{\Omega_{p2}}+ ({W_{22}}{\left| {{\Omega_{p2}}} \right|^2} + {W_{21}}{\left| {{\Omega_{p1}}} \right|^2}){\Omega_{p2}} \notag \\
&+ \int {{d^3}{ r'}{\mathcal{N}_{22}({\bf r'} -{\bf  r})}{{\left| {{\Omega_{p2}({\bf r'})}} \right|}^2} {\Omega_{p2}({\bf r})}}\notag\\
&+  \int {{d^3}{ r'}{\mathcal{N}_{21}({\bf r'} -{\bf  r})}{{\left| {{\Omega_{p1}({\bf r'})}} \right|}^2}{\Omega_{p2}({\bf r})}}  = 0,
\end{align}
\end{subequations}}
where
{\small
\begin{subequations}\label{coefWs}
\begin{align}
&{W_{1l}} = \kappa_{13}\big[\Omega _c^*a_{43,l}^{(2)} + {d_{41}}(a_{11,l}^{(2)} - a_{33,l}^{(2)})+\delta_{l2}a_{21}^{(2)}\big]/D_1, \\
&{W_{2l}} = \kappa_{23}\big[\Omega _c^*a_{43,l}^{(2)} +  {d_{42}}(a_{22,l}^{(2)} - a_{33,l}^{(2)}) + \delta_{l2}a_{21}^{*(2)}\big]/D_2, \\
&\mathcal{N}_{jl} = \frac{1}{2}\kappa_{13}\Omega _c^*{\mathcal{N}_a}\mathcal{V}({\bf r'} -{\bf  r})a_{44,4j,l}^{(3)}({\bf r'} -{\bf  r})/D_j.
\end{align}
\end{subequations}}\noindent
The coefficient $W_{jl}$~[$\mathcal{N}_{jl}({\bf r'} -{\bf  r})]$ characterizes the local~(nonlocal) self-Kerr~($j=l$)  and cross-Kerr~($j\neq l$) nonlinearity of the $j$th polarization components of the probe field.  For simplicity, we assume that the probe field is slowly varied along the $z$ direction, so that a local approximation in this direction can be made for the nonlinear response function. Then Eqs.~(\ref{cnls}) are reduced to
{\small \begin{subequations}\label{cnls20}
\begin{align}
i\frac{\partial \Omega_{p1}}{\partial z}& + \frac{c}{2\omega _{p}}\nabla _ \bot ^2\Omega_{p1}+ (W_{11}\left| \Omega_{p1} \right|^2 + W_{12}\left| \Omega_{p2}\right|^2)\Omega_{p1} \notag \\
&+ \int d^2 r' \big[\mathcal{N}_{11}'({\bf r}_\bot' -{\bf  r}_\bot)\left| \Omega_{p1}({\bf r}_\bot',z) \right|^2 \notag\\
&+\mathcal{N}_{12}'({\bf r}_\bot' -{\bf  r}_\bot)\left| \Omega_{p2}({\bf r}_\bot',z) \right|^2\big]\Omega_{p1}({\bf r}_\bot) = 0, \\
i\frac{{\partial {\Omega_{p2}}}}{{\partial z}} & + \frac{c}{{2{\omega _{p}}}}\nabla _ \bot ^2{\Omega_{p2}}+ ({W_{22}}{\left| {{\Omega_{p2}}} \right|^2} + {W_{21}}{\left| {{\Omega_{p1}}} \right|^2}){\Omega_{p2}} \notag \\
&+ \int {d^2}{r_\bot'}\big[{\mathcal{N}_{22}'({\bf r}_\bot' -{\bf  r}_\bot)}{{\left| {{\Omega_{p2}({\bf r}_\bot',z)}} \right|}^2}\notag\\
& + {\mathcal{N}_{21}'({\bf r}_\bot' -{\bf  r}_\bot)}{{\left| {{\Omega_{p1}({\bf r}_\bot',z)}} \right|}^2}\big]{\Omega_{p2}({\bf r}_\bot)}  = 0,
\end{align}
\end{subequations}}\noindent
where ${\bf r}_\bot=(x,y)$ and $d^2r_\bot=dx'dy'$,  the reduced nonlocal nonlinear response function reads $\mathcal{N}'_{jl}({\bf r}_\bot)=\int  \mathcal{N}_{jl}({\bf r})dz$.
Equations (\ref{cnls20}) can be cast into the dimensionless form
{\small \begin{align}
&  i\frac{\partial u_1}{\partial s}+\tilde{\nabla}_\bot^2u_1+(w_{11}|u_1|^2+w_{12}|u_2|^2)u_1
\notag\\
&+\int d^2\zeta'\big[\Re_{11}(\vec{\zeta}'-\vec{\zeta})|u_1(\vec{\zeta}')|^2+\Re_{12}(\vec{\zeta}'-\vec{\zeta})|u_2(\vec{\zeta}')|^2\big]u_1=0,\\
&  i\frac{\partial u_2}{\partial s}+\tilde{\nabla}_\bot^2u_2+(w_{22}|u_2|^2+w_{21}|u_1|^2)u_2
\notag\\
&+\int d^2\zeta'\big[\Re_{22}(\vec{\zeta}'-\vec{\zeta})|u_2(\vec{\zeta}')|^2+\Re_{21}(\vec{\zeta}'-\vec{\zeta})|u_1(\vec{\zeta}')|^2\big]u_2=0,
\end{align}}\noindent
where $u_j=\Omega_{pj}/U_0$~($U_0$ is the typical half Rabi frequency of the probe field), $s=z/(2L_{\rm diff})$~($L_{\rm diff}=\omega_pR_0^2/c$ is the typical diffraction length; $R_0$ the typical transverse radius of the probe beam), $w_{jl}=2U_0^2L_{\rm diff} W_{jl}$~($jl=\{11,22,12,21\}$) are the dimensionless local  nonlinear coefficients, $\Re_{jl}(\vec{\zeta}'-\vec{\zeta})=2L_{\rm diff}R_0^2U_0^2\mathcal{N}'_{jl}[(\vec{\zeta}'-\vec{\zeta})R_0]$, $\tilde{\nabla}_\bot^2=\partial^2/\partial\xi^2+\partial^2\partial\eta^2$, $\vec{\zeta}=(\xi,\eta)=(x,y)/R_0$, and $d^2\zeta'=d\xi'd\eta'$.

\section{A short description of numerics details}\label{appD}

In this section, we give a simple description on the imaginary-time propagation method~\cite{YangJK2010} for
numerically finding the steady-state solutions (including stationary optical patterns, solitons and vortices) of the dimensionless nonlinear Schr\"odinger equation (\ref{cnls3}) used in the main text.  The basic ideas are the following: (i)~Replacing $s$ by $-is$ in the equation;  (ii)~Normalizing the solution after each step of integration, i.e., $v(\xi,\eta,s+\Delta s)=\frac{v(\xi,\eta,s+\Delta s)}{\|v(\xi,\eta,s+\Delta s)\|}$ to keep the power of the solution to a fixed value.

Based on these ideas, firstly we replace $s\rightarrow -is$ in Eq.~(\ref{cnls3}), which results in the new equation
{\small\begin{align}\label{imcnl}
&\frac{\partial v_j}{\partial s}=\tilde{\nabla}_\bot^2v_j+\sum_{l=1,2}\int d^2\zeta'\big[I_l\Re_{jl}(\vec{\zeta}'-\vec{\zeta})|v_l(\vec{\zeta}',s)|^2\big]v_j(\vec{\zeta},s).
\end{align}}\noindent
Secondly, we solve the above equation (\ref{imcnl}) by a split-step Fourier method.  The first step is to integrate the diffraction term in momentum space by a fast Fourier transform, and then go back to the spatial coordinate by an inverse Fourier transform; the second step is to carry out the integration on the terms with nonlocal interactions, which is solved by a numerical convolution. Thus we have
\begin{subequations}
  \begin{align}
  &\frac{d \tilde{v}_j}{ds}=-\beta^2\tilde{v}_j,\label{cc1}\\
  &\frac{d v_j}{ds}=\sum_{l=1,2}I_l\mathcal{F}^{-1}\Big\{\mathcal{F}\big[\Re_{jl}(\vec{\zeta})\big]\mathcal{F}
  \big[|v_l(\vec{\zeta},s)|^2\big]\Big\}v_j(\vec{\zeta},s),\label{cc2}
\end{align}
\end{subequations}
where $-\beta^2$ and $\tilde{v}_j$  are respectively the Fourier transforms of $\tilde{\nabla}_\bot^2$ and $v_j$;  $\mathcal{F}$ and $\mathcal{F}^{-1}$ are respectively the symbols of the Fourier transform and  the inverse Fourier transform.

The imaginary-time propagation is made in square domain until the convergence of the intensity distribution $|v_j|^2$ within the error less than $10^{-6}$. In the simulation, the initial condition is a plane wave perturbed by Gaussian noise; the boundary condition is a periodic one.  The simulations are implemented by using different numerical grids~(i.e., 256, 512, and 1024) in the transverse (i.e. $\xi$ and $\eta$) directions, which can ensure the results (including the ones for the MI with short wavelengths) for the different grids to be coincident with each other.

\section{Details for the alternative emergence of the circular and elliptical spots in Fig.~\ref{fig4}(a)}\label{app3}

Here we give a detailed illustration on the alternative emergence of circular and elliptical spots shown in Fig.~\ref{fig4}(a). This steady-state pattern is obtained by numerically solving Eq.~(\ref{cnls3}) by using imaginary propagation together with split-step Fourier methods, by taking the system parameters $(\alpha_{12}/\alpha_{11},\alpha_{11})=(0.03, 245)$,  $\sigma=1$, and $\rho_{22}^{(0)}/\rho_{11}^{(0)}=1$.

\begin{figure}[htpb]
\centering
\includegraphics[width=0.5\textwidth]{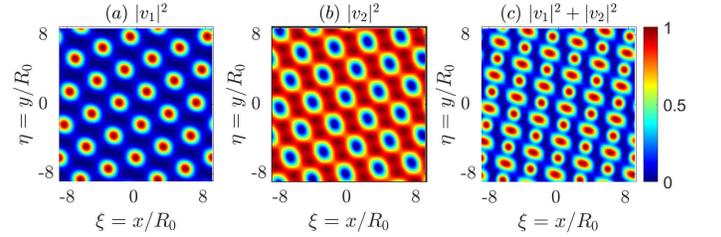}
\caption{\footnotesize Optical patterns of the two-component probe field as functions of $\xi = x/R_0$ and $\eta = y/R_0$, by taking $(\alpha_{12}/\alpha_{11},\alpha_{11})=(0.03, 245)$, $\sigma=1$, and $\rho_{22}^{(0)}/\rho_{11}^{(0)}=1$. (a)~The result for $|v_1|^2$, which is a hexagonal lattice with circular bright spots in a dark background. (b)~The result for $|v_2|^2$, which is also a hexagonal lattice but with circular bright spots in a bright background. (c)~The result for $|v|^2=|v_1|^2+|v_2|^2$, which is a linear superposition of the two polarized components, giving a pattern where the circular and elliptical spots appear alternatively.
}\label{fig9}
\end{figure}

Shown in the panels (a), (b), and (c) of the Fig.~\ref{fig9} are results of dimensionless and normalized probe-field intensities $|v_1|^2$, $|v_2|^2$, and $|v|^2=|v_1|^2+|v_2|^2$ as functions of the dimensionless coordinates $\xi = x/R_0$ and $\eta = y/R_0$, respectively.
We see that both the light intensities $|v_1|^2$ and $|v_2|^2$ of the two probe-field components form hexagonal lattices, given respectively by Fig.~\ref{fig9}(a) and Fig.~\ref{fig9}(b). However, their superposition,  $|v_1|^2+|v_2|^2$, gives a lattice pattern in which the $v_1$-component looks like to be composed of circular spots but  the $v_2$-component looks like to be composed of elliptical spots [see Fig.~\ref{fig9}(c)]. The reason for the appearance of such an interesting pattern is that the $v_1$-component has a hexagonal structure composed of circular bright spots (where light intensity is maximal) in a dark background, while the $v_2$-component has a hexagonal structure composed of circular dark spots (where light intensity is minimal) in a bright background. As a result, the intensity superposition of the two polarized components (i.e., $|v_1|^2+|v_2|^2$) gives a pattern in which the circular and elliptical spots appear alternatively. A similar phenomenon also occurs in Fig.~\ref{fig5}(a), with the physical reason the same as given above.

\section{Derivation of the criterion for the optical phase separation}\label{appC}

We follow the method used in Refs.~\cite{Ao1998,Kumar2019,Wen2012} to find the criterion for the transition between miscible and immiscible states through minimizing the energy of the system.
For the miscible state, the distribution of $v_j=v_j(\vec{\zeta},s)$  is in the whole space. Using Eq.~(\ref{energy}), we have
{\small \begin{align}
&E_{\rm mis}=\sum_{j=1}^2\int |\tilde{\nabla}_{\perp} v_j|^2d^2\zeta+\frac{1}{2}\sum_{j,l=1}^2E_{jl},
\end{align}}\noindent
where $E_{jl}=\iint \Re_{jl}(\vec{\zeta}'-\vec{\zeta})|v_j(\vec{\zeta},s)|^2|v_l(\vec{\zeta}',s)|^2d^2\zeta d^2\zeta'$.
For simplicity, we consider a homogeneous
solution of the Eqs.~(\ref{cnls3}). The corresponding energy reads
{\small \begin{align}\label{energy3}
E_{\rm mis} &=\frac{1}{2V_{\rm tot}}\int\big\{I_1^2\mathcal{\Re}_{11}(\vec{\zeta})+ I_2^2 \mathcal{\Re}_{22}(\vec{\zeta})\notag\\
&+I_1I_2[\mathcal{\Re}_{12}(\vec{\zeta})+\mathcal{\Re}_{21}(\vec{\zeta})]\big\}d^2\zeta,
\end{align}}\noindent
where $V_{\rm tot}=d^2\zeta$ is the volume of the whole space.  For immiscible state,
the two components occupy different positions in space. Assume
$V_1$ ($V_2$) is the volume occupied by the component $1$ (component $2$), the overlap  integrals in Eq.~(\ref{energy3}) will be zero. So the energy of the immiscible state is given by
{\small \begin{align}\label{energy2}
 &E_{\rm immis}= \sum_{j=1}^2\int |\tilde{\nabla}_\perp v_j|^2d^2\zeta+\frac{1}{2}\sum_{j=1}^2E_j,
\end{align}}\noindent
where $E_j=\iint \Re_{jj}(\vec{\zeta}'-\vec{\zeta})|v_j(\vec{\zeta},s)|^2|v_j(\vec{\zeta}',s)|^2d^2\zeta d^2\zeta'$. The integral for $\mathcal{\Re}_{jj}$ is for the component $j$ which occupies the volume $V_j$, with $V_1$ and $V_2$
satisfying $V_{\rm tot}=V_1+V_2$. We obtain
{\small \begin{align}
E_{\rm immis}=&\frac{1}{2V_{\rm tot}}\Big[\int\big\{I_1^2\mathcal{\Re}_{11}(\vec{\zeta})
+ I_2^2 \mathcal{\Re}_{22}(\vec{\zeta})\big\}d^2\zeta\notag\\
&+I_1I_2\Big(\int \mathcal{\Re}_{11}(\vec{\zeta})d^2\zeta
\times\int \mathcal{\Re}_{22}(\vec{\zeta})d^2\zeta\Big)^{1/2}\Big].
\end{align}}\noindent

 The energy difference between the miscible and immiscible state  is
 {\small \begin{align}\label{MIT}
 \Delta E=& \frac{I_1I_2}{V_{\rm tot}} \Big[
\int \frac{1}{2}\big[\mathcal{\Re}_{12}(\vec{\zeta})+\mathcal{\Re}_{21}(\vec{\zeta})\big] d^2\zeta\notag\\
&-\Big(\int \mathcal{\Re}_{11}(\vec{\zeta})d^2\zeta \times\int \mathcal{\Re}_{22}(\vec{\zeta})d^2\zeta\Big)^{1/2}\Big].
\end{align}}\noindent
Therefore, $\Delta E =0$ defines the critical value (boundary) for the transition from miscible  to immiscible  systems.



\begin{references}
\bibitem{note0}
We assume all the atoms are initially populate in the states $|1\rangle$ and $|2\rangle$ equally, and hence the average of any atomic operator $\hat{O}$ amounts to $\langle \hat{O}\rangle=\langle G|\hat{O}|G\rangle$, with $|G\rangle=|1_1,1_2,,...,1_{N/2},2_{N/2+1},2_{N/2+2},...,2_N\rangle$.
\bibitem{note99}
The nonlinearity length of the system is given by $L_{\rm non}\equiv [U_0^2 \int d^2 r_{\perp}^{\prime} |N_{11}({\bf r}_{\perp}^{\prime})|]^{-1}$. With this definition,
one has $\Re_{jl}(\vec{\zeta}'-\vec{\zeta})=2(L_{\rm diff}/L_{\rm non})\,B_0 N_{jl}^{\prime}[(\vec{\zeta}'-\vec{\zeta})R_0]$, with $B_0\equiv R_0^2/\int d^2 r_{\perp} |N_{11}({\bf r}_{\perp})|$.
\bibitem{note100}
In our consideration, $U_0=6.28$\,MHz, $R_0\approx 7\,\mu$m, and $L_{\rm diff}\approx L_{\rm non}= 0.2$\,mm.
\bibitem{note1}
Generally, the growth rate is a complex number, i.e. $\lambda=\lambda_r+i\lambda_i$. We find that $\lambda$ is either a real number ($\lambda=\lambda_r$) or an imaginary one ($\lambda=i\lambda_i$). Thus in linear level either the perturabtrion  grows exponentially when $\lambda^2=\lambda_r^2>0$, or it decays exponentially when $\lambda^2=-\lambda_i^2<0$.
\bibitem{note2}
Note that in our model the two orthogonal, circularly polarized components ($\sigma^+$ and $\sigma^-$) of the probe field couple the atomic transitions  $|1\rangle\leftrightarrow |3\rangle$ and $|2\rangle\leftrightarrow |3\rangle$, respectively. Hence the normalized total probe-field intensity is given by $I=|v_1|^2+|v_2|^2$.
\bibitem{note3}
In particular, if $\Re_{ij}(\vec{\zeta})=g_{ij}\delta(\vec{\zeta})$ the definition (\ref{MIT0}) is simplified to $\Delta E=\frac{I_1I_2}{V_{\rm tot}} \Big[ \frac{1}{2}(g_{12}+g_{21})-\sqrt{g_{11}g_{22}}\Big]$, which has been used to characterize the phase separation in systems with local
nonlinearities (i.e. $\sigma=0$)~\cite{Ao1998,Kumar2019,Wen2012}. In this case, the critical value for the transition from the miscibility to the immiscibility is given by $\alpha_{12}/\alpha_{11}=1$.
\end{references}
\end{document}